\documentclass[%
 aps,
 prb,%
 amsmath,amssymb,
 reprint,%
superscriptaddress,
]{revtex4-1}

\usepackage{graphicx}
\usepackage{dcolumn}
\usepackage{bm}
\usepackage{color}
\usepackage{longtable}
\usepackage[colorlinks=true,allcolors=blue]{hyperref}

\begin{document}
\newcommand{\micron}{$\mu$m}

\title{Benchmarking boron carbide equation of state using computation and experiment}
\author{Shuai Zhang}
\email{szha@lle.rochester.edu}
\affiliation{Lawrence Livermore National Laboratory, Livermore, California 94550, USA}
\affiliation{Laboratory for Laser Energetics, University of Rochester, Rochester, New York 14623, USA}
\author{Michelle C. Marshall}
\email{marshall47@llnl.gov}
\author{Lin H. Yang}
\author{Philip A. Sterne}
\affiliation{Lawrence Livermore National Laboratory, Livermore, California 94550, USA}
\author{Burkhard Militzer}
\email{militzer@berkeley.edu}
\affiliation{Department of Earth and Planetary Science, University of California, Berkeley, California 94720, USA}
\affiliation{Department of Astronomy, University of California, Berkeley, California 94720, USA}
\author{Markus D\"ane}
\author{James A. Gaffney}
\author{Andrew Shamp}
\author{Tadashi Ogitsu}
\author{Kyle Caspersen}
\affiliation{Lawrence Livermore National Laboratory, Livermore, California 94550, USA}
\author{Amy E. Lazicki}
\author{David Erskine}
\author{Richard A. London}
\author{Peter M. Celliers}
\author{Joseph Nilsen}
\author{Heather D. Whitley}
\email{whitley3@llnl.gov}
\affiliation{Lawrence Livermore National Laboratory, Livermore, California 94550, USA}

\date{\today}
{\begin{abstract}
{
Boron carbide (B$_4$C) is of both fundamental scientific and practical interest due to its structural complexity and how it changes upon compression, as well as its many industrial uses and potential for use in inertial confinement fusion (ICF) and high energy density physics experiments.
We report the results of a comprehensive computational study of the equation of state (EOS) of B$_4$C in the liquid, warm dense matter, and plasma phases.
Our calculations are cross-validated by comparisons with Hugoniot measurements up to 61 megabar from planar shock experiments performed at the National Ignition Facility (NIF).
Our computational methods include path integral Monte Carlo, activity expansion, as well as all-electron Green's function Korringa-Kohn-Rostoker and molecular dynamics that are both based on density functional theory.
We calculate the pressure-internal energy EOS of B$_4$C over a broad range of temperatures ($\sim$6$\times$10$^3$--5$\times$10$^8$~K) 
and densities (0.025--50 g/cm$^{3}$).
We assess that the largest discrepancies between theoretical predictions are $\lesssim$5\% near the compression maximum at 1--2$\times10^6$ K. This is the warm-dense state in which the K shell significantly ionizes and has posed grand challenges to theory and experiment.
By comparing with different EOS models, we find a Purgatorio model (LEOS 2122) that agrees with our calculations.
The maximum discrepancies in pressure between our first-principles predictions and LEOS 2122 are $\sim$18\% and occur at temperatures between 6$\times$10$^3$--2$\times$10$^5$~K,
which we believe originate from differences in 
the ion thermal term and 
the cold curve that are modeled in LEOS 2122 in comparison with our first-principles calculations.
In order to account for potential differences in the ion thermal term, we have developed three new equation of state models that are consistent with theoretical calculations and experiment.  We apply these new models to 1D hydrodynamic simulations of a polar direct-drive NIF implosion, demonstrating that these new models are now available for future ICF design studies. (LLNL-JRNL-812984)
}
\end{abstract}
}


\maketitle

\section{Introduction}\label{sec:introd}

The design of high energy density and inertial confinement fusion experiments
requires a good description of the ablator equation of state (EOS).
Materials that are typically used as ablators are plastics, such as
hydrocarbons (CH) and glow discharge polymers (GDP).~\cite{Barrios2010CH,Barrios2012GDP,Doppner2018prl-CH,Kritcher2016}
However, formation of condensed phase microstructures and mixing
with the DT fuel during implosion could affect the performance 
of the ignition target~\cite{Clark2016,Guskov2017}.
Additional materials with higher density and hardness,
such as high-density carbon (HDC), boron-materials, and beryllium also provide current and future
options for ablators.~\cite{Moore2016b,OLSON2007,MacKinnon2014,Kritcher2018,Zylstra2019}.
In comparison to plastics, these high-tensile strength materials
typically exhibit ablation pressures that are 
15-20\% higher~\cite{Moore2016b}.
Using these materials as the ablator can have higher x-ray absorption
and use a shorter laser pulse with a
higher ablation rate for a given temperature,
and thereby require a thinner ablator shell while 
maintaining the same mass and outer diameter~\cite{Moore2016b,Kritcher2018,Zhang2018b}. 
Ablators doped with boron have also been the subject of more
recent proposals to use reactions with $\gamma$-rays as a means of quantifying ablator mix in inertial confinement fusion (ICF) experiments,\cite{Hayes17}
and boron carbide is of particular interest for ignition experiments because a method for producing hollow capsules has already been demonstrated.\cite{CHEN2017571}

In recent studies, Zhang {\it et al.} combined several computational methods
to set accurate constraint for the EOS 
of boron (B)~\cite{Zhang2018b} and boron nitride (BN)~\cite{Zhang2019bn}
over a wide range of temperatures ($\sim$0.2 eV--50 keV) and 
densities (0.1--20 times compression).
They also conducted laser shock experiments 
at the Omega laser facility and the National Ignition Facility (NIF) 
to measure
the Hugoniot EOS to pressures of 10--60 megabar (Mbar)
and demonstrated remarkable agreement with the 
first-principles predictions.
Their data have enabled building new EOS tables 
(X52 for B and X2152 for BN) based on the 
quotidian EOS (QEOS) model~\cite{leos1qeos,leos2}
and clarifying the dominating physics 
(cold curve, ion thermal, or electron thermal) 
at different regions of the temperature-density space.
They also performed 1D hydrodynamic simulations of polar
direct-drive exploding-pusher experiments~\cite{Ellison_2018}
to explore the performance sensitivity to the EOS.

Boron carbide is another important member in the family of 
boron materials. At ambient condition, it has a high melting point, 
superior hardness, low specific weight, good resistance to chemical
agents, and high neutron absorption cross section. 
These outstanding properties allow it to be widely 
used for mechanical, electrical, chemical, and
nuclear applications.~\cite{THEVENOT1990205}
The ambient crystal structure of B$_4$C has rhombohedral
symmetry (space group R{\=3}m), similar to that of $\alpha$-B,
and is characterized by B-rich icosahedra and C-rich chains.
X-ray diffraction experiments reveal this structure to be
stable under static compression at up to 126 GPa.~\cite{Fujii_2010}
Single-crystal experiments show that the icosahedral units
are less compressible than the unit cell volume and the static compression
is governed by force transfer between the rigid icosahedra.~\cite{DERA201485}
However, dramatic structural changes have been reported under
shock compression~\cite{Chen2003b4c,Zhang2006,Volgler2004}, 
scratching and nanoindentation~\cite{Domnich2002,GE20043921,Yan2006}, 
or depressurization~\cite{Yan2009} and attributed to amorphization or 
structural transition that is accompanied by changes in hardness, compressibility, 
or elastic modulus.~\cite{Grady2015,Chen2003b4c,Awasthi2019}
There have also been studies that show the shear strength of boron-rich 
boron carbide can be lowered due to nanotwins~\cite{An2017bc}
and multi-scale molecular dynamics (MD) simulations that relate
structural changes to hydrostaticity of compression~\cite{Korotaev2017}.

Over the last few years, knowledge about the EOS of boron carbide
has advanced significantly. The EOS and melting curve of B$_4$C were constructed by
Molodets {\it et al.}~\cite{Molodets2017} that agree with available 
experiments at up to megabar pressures, featuring melting with
a negative Clapeyron slope at pressures below 150 GPa
and a positive one above 170 GPa.
Jay {\it et al.}~\cite{Jay2019} performed comprehensive {\it ab initio} 
calculations for boron carbide at up to 80~GPa and 2000~K, and their
temperature-pressure-concentration phase diagrams show phase separation 
of boron carbides in multiple stages and into B and C at above 70 GPa.
Fratanduono {\it et al.}~\cite{Fratanduono_B4C} extended the Hugoniot,
sound velocities, and thermodynamic properties measurements of liquid B$_4$C 
to 700 GPa. Shamp {\it et al.}~\cite{Shamp2017} performed MD calculations
based on density functional theory (DFT) to determine the Hugoniot curve up to
1500 GPa, and predicted discontinuities along the Hugoniot at $<$100 GPa
as results of phase separation and transformation in solid B$_4$C. 
An equation of state table (LEOS 2122) based on an average atom-in-jellium model (Purgatorio)~\cite{Purgatorio2006} 
has thus been developed that fits all available 
experimental Hugoniot data above 100 GPa~\cite{Fratanduono_B4C}. 
However, accurate EOS at higher pressures and temperatures, 
in particular those corresponding to the partially ionized, warm dense state, 
is still unknown.

The goal of this work is to benchmark the EOS of B$_4$C in a wide range of 
temperatures and pressures by combining theoretical calculations and experiments.
Our theoretical methods include path integral Monte Carlo (PIMC), 
pseudopotential DFT-MD approaches realized in multiple schemes, 
an activity expansion method (ACTEX), and an all-electron, 
Green's function Korringa-Kohn-Rostoker (KKR) method. 
Our experiments consist of seven Hugoniot measurements conducted 
at the NIF. The paper is organized as follows: 
Sec.~\ref{sec:theorymethod} outlines our computational details; 
Sec.~\ref{sec:expt} describes our shock experiments; 
Sec.~\ref{sec:results} compares our EOS and Hugoniot results from computation 
and experiments, constructs new EOS models, and explores the role of EOS in hydrodynamic simulations;
Sec.~\ref{sec:discuss} discusses the microscopic physics of B$_4$C 
by combining electronic structure and QEOS perspectives;
finally we conclude in Sec.~\ref{sec:conclusion}.

\section{Computational methods}\label{sec:theorymethod}

\begin{figure}
\centering\includegraphics[width=0.48\textwidth]{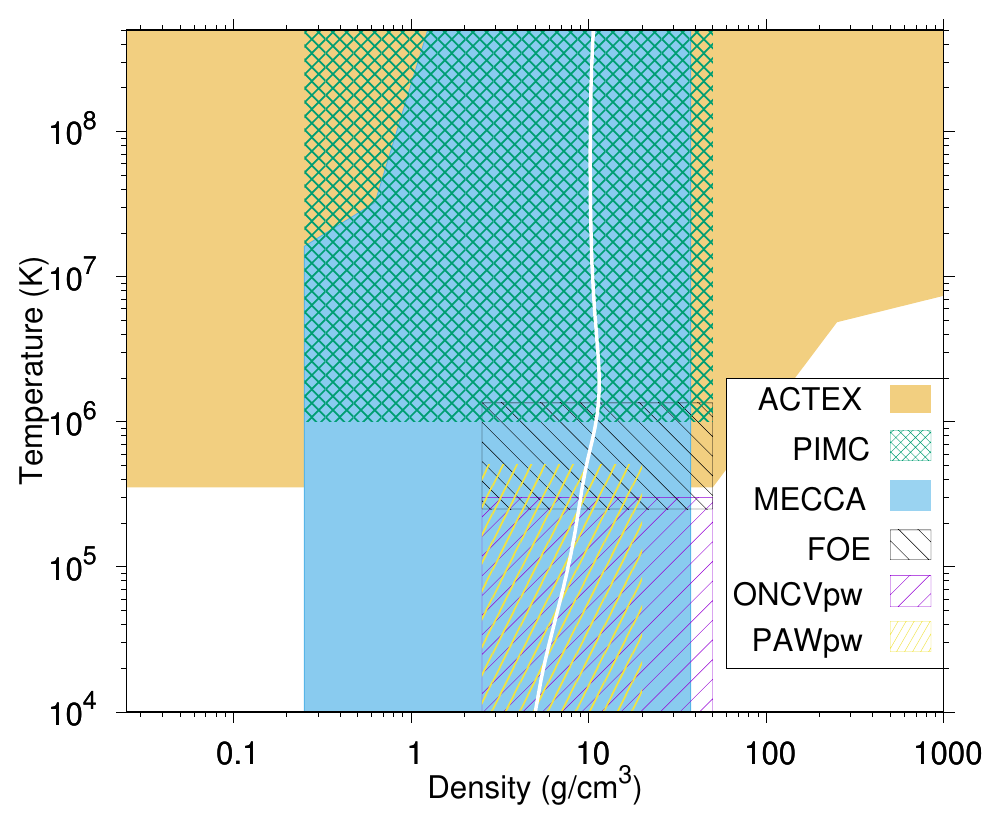}
\caption{\label{fig:theorymethods} Schematic diagram showing the temperature-density regions at which different methods are used in this work for calculating the EOS of B$_4$C. The principle Hugoniot from LEOS 2122 is shown (white curve) for comparison.}
\end{figure}

In this section, we briefly describe the computational settings of 
the theoretical methods that we employ to compute the internal energies 
and pressures of B$_4$C across wide ranges of temperatures and densities. 
Figure~\ref{fig:theorymethods} summarizes the conditions at which 
each of the methods has been used.
The computations are performed by leveraging the applicability,
accuracy, and efficiency of each method. 
More theoretical details can be found in our recent 
paper~\cite{Zhang2019bn} and references therein. 

We perform PIMC simulations of B$_4$C using the {\footnotesize CUPID} code~\cite{militzerphd}.
All electrons and nuclei are treated explicitly. In order to deal with 
the Fermionic sign problem, we apply the fixed-node approximation
using free-particle nodes to restrict the paths~\cite{ceperley1995,Pierleoni1994,Driver2012}.
The pair density matrices~\cite{Na95,pdm} are evaluated in steps 
of $\frac{1}{512}$~Hartree$^{-1}$ (Ha$^{-1}$) and the
nodal restriction is enforced in steps of 
$\frac{1}{8192} ~\text{Ha}^{-1}$.
The calculations are performed at densities of 
0.25--50.17~g/cm$^3$ [0.1 to 20 times the ambient density ($\rho_0\sim2.5$~g/cm$^3$)~\cite{Clark1943B4C}]
and temperatures of $10^6$--5$\times$10$^8$~K.
Each simulation cell consists of 30 atoms, which is comparable to our previous simulations 
for pure B, BN, and hydrocarbons~\cite{Zhang2018b,Zhang2019bn,Zhang2017b,Zhang2018}. 
The finite cell size effects on the EOS are negligible 
at such high temperature conditions~\footnote{By comparing the EOS and 
the radial distribution function $g(r)$ obtained using 30-atom cells to 
those using 120-atom cells in our DFT-MD calculations, we find negligible 
differences at temperatures above $5\times10^4$ K. 
A comparison in $g(r)$ is shown in Fig.~\ref{fig:gr}.}.


Our DFT-MD simulations for B$_4$C are performed in two different ways. 
One way is by using the frozen-1s-core projector augmented wave (PAW)~\cite{Blochl1994} 
or optimized norm-conserving Vanderbilt (ONCV) pseudopotentials~\cite{oncv13,oncv13e}
and plane-wave (pw) basis;
the other is a Fermi operator expansion (FOE)~\cite{Goedecker99,Bowler2012}
approach using all-electron ONCV potentials.
The PAWpw calculations are performed using the
Vienna \textit{Ab initio} Simulation Package
({\footnotesize VASP})~\cite{kresse96b}
and employing the hardest available PAW potentials (core radius 
equals 1.1 Bohr for both B and C),
Perdew-Burke-Ernzerhof (PBE)~\cite{Perdew96} 
exchange-correlation functional, 
a large cutoff energy (2000~eV) for the plane-wave basis, 
and the $\Gamma$ point to sample the Brillouin zone.
The PAWpw calculations of the EOS are performed at 
6.7$\times$10$^3$--5.05$\times 10^5$~K ($\sim$0.6--43.5~eV)
and 1--10 times $\rho_0$.
We conducted ONCVpw simulations~\cite{Yang2007}
at temperatures up to 3$\times10^5$ K, 
using PBE exchange-correlation functional and a 900 eV energy cutoff (core radius equals 1.125 Bohr for both B and C) for the plan-wave expansion, in order
to cross check the PAWpw results.
For both PAWpw and ONCVpw calculations, 
a Nos\'{e} thermostat~\cite{Nose1984} is used to generate 
MD trajectories (typically $\sim$5000 steps) that form 
canonical ensembles.
The MD time step is chosen within the range of 0.05-0.55~fs, 
smaller at higher temperatures.
Cubic cells with 30 and 120 atoms are considered to eliminate
the finite-size errors on the EOS.


We perform FOE calculations at temperatures of 
$2.5\times 10^5$--$1.34\times 10^6$~K.
Note that FOE takes advantage of the smooth 
Fermi-Dirac function at high temperature by 
approximating the function with polynomial expansion, 
which provides a very efficient way to conduct the 
Kohn-Sham DFT-MD calculation.
We use 30-atom cells
and conduct $NVT$ simulations that last
3000--6000 steps 
(0.05--0.1~fs/step) to ensure sufficient statistics
to obtain the EOS. To be consistent with the plan-wave calculations, the FOE calculations employ 
PBE exchange-correlation functional and much larger energy cutoff (4000 eV) due to smaller core radius (0.8 Bohr) due to the inclusion of 1s core states in both B and C pseudopotentials. 
We also use the all-electron ONCV potentials and pw basis to perform calculations at densities of 12.544 g/cm$^3$ or higher and temperatures of $1.26\times10^5$ K or lower, in order to reduce the possibility of frozen-core overlap in the MD simulations.

Over the last ten years, Militzer {\it et al.} have developed 
and employed the approach combining PIMC and DFT-MD to calculate
the EOS of a series of elemental materials (He~\cite{Militzer2009He}, B~\cite{Zhang2018b}, C~\cite{Driver2012}, N~\cite{Driver2016Nitrogen}, O~\cite{Driver2015Oxygen}, Ne~\cite{Driver2016Nitrogen}, Na~\cite{Zhang2016b,Zhang2017}, Al~\cite{Driver2018Al}, Si~\cite{Militzer2015})
and compounds (H$_2$O~\cite{Driver2012},LiF~\cite{Driver2017LiF},CH~\cite{Zhang2017b,Zhang2018}, BN~\cite{Zhang2019bn}, MgO~\cite{Soubiran2019}, MgSiO$_3$~\cite{Gonzalez-Cataldo2020})
over wide ranges of temperatures and pressures.
The PIMC data were shown to reproduce predictions by classical
plasma theories (such as the Debye-H\"uckel and 
the ideal Fermi-gas model)
in the limit of infinitely high temperatures and agree remarkably 
well (differences up to $\sim5$\%) with
DFT-MD for the partially ionized, warm dense states at 
$\sim10^5$--$10^6$~K (or 10--100 eV), while the DFT-MD predictions
of the Hugoniot are consistent with dynamic shock experiments
that are available up to multi-megabar (Mbar) pressures.
By fully capturing the ionic interaction effects (DFT-MD),
nuclear quantum effects (PIMC), and electronic many-body effects (PIMC),
these computations set accurate constraints for the EOS of these
materials ($Z$ up to 14) from condensed matter to hot plasma 
states (degeneracy parameter $\sim$0.1--$10^3$, coupling parameter
$\sim$0.01--10) and serve as benchmarks for the development of other,
computationally more efficient EOS methods.

In a recent paper~\cite{Zhang2019bn}, an all-electron, 
Green's function KKR electronic-structure method 
based on Kohn-Sham DFT and an activity expansion
method,
in addition to FOE and a spectral quadrature method,
were used to compute the EOS of BN
and compare with the PIMC and pw DFT-MD data.
The Green's function method simplifies the calculation
by using a static lattice and 
approximating the ion kinetic contribution with
an ideal gas model, and show good agreement with
PIMC and DFT-MD predictions at above $10^5$ K
when the ion thermal contribution becomes less 
significant in comparison to electron thermal
or cold curve contributions.
The activity expansion approach is based on an expansion
of the plasma grand partition function in powers of the 
constituent particle activities (fugacities)~\cite{rogers73,rogers74},
and the EOS calculations include interaction terms beyond the
Debye-H\"uckel, electron-ion bound states and ion-core plasma 
polarization terms, along with relativistic and quantum 
corrections~\cite{rogers79,rogers81}, and therefore 
produce accurate EOS at temperatures down to $\sim$10$^6$~K.
It is thus interesting to explore the ranges of applicability
of these approaches for B$_4$C.

We use the Multiple-scattering Electronic-structure Calculation for 
Complex Applications ({\footnotesize MECCA}) code for the all-electron, Green's function KKR calculations.~\cite{MECCA-2015} 
The KKR spherical-harmonic local basis included 
$L_\text{max}=2$ within the multiple-scattering contributions, 
and $L$ up to 200 are included automatically 
until the free-electron Bessel functions contribute zero 
to the single-site wavefunction normalizations.
We use local density approximation (LDA)~\cite{doi:10.1139/p80-159}
for the exchange-correlation functional,
a 12$\times$12$\times$12 Monkhorst-Pack~\cite{Monkhorst1977} $k$-point mesh
for Brillouin zone integrations for energies with an imaginary part smaller 
than 0.25~Rydberg, and a 8$\times$8$\times$8 $k$-point mesh otherwise.
A denser mesh was used for the physical density of states calculated 
along the real-energy axes when needed. 
We use a static 5-atom cubic cell for the calculations and approximate the ion-kinetic contribution by the ideal gas model. This structure can be viewed as a body-centered cubic carbon lattice that has a simple-cubic boron sublattice inscribed at $(\pm1/4,\pm1/4,\pm1/4)$ and $(\pm3/4,\pm3/4,\pm3/4)$. This assumed crystal structure is by no means representable for B$_4$C at ambient conditions. Therefore, it is not expected to agree with experiments or other computational methods that do not assume this static structure. However, the structure is space filling and might be a representation for higher temperatures and pressures. 

Activity expansion calculations are performed using 
the {\footnotesize ACTEX} code~\cite{rogers73,rogers74}.
We cut off {\footnotesize ACTEX} calculations at temperatures 
below the point where many-body terms become comparable to the
leading-order Saha term ($T>5.8\times10^5$~K). 

\section{Experiments}\label{sec:expt}

\begin{figure*}
	\includegraphics[width=0.6\textwidth]{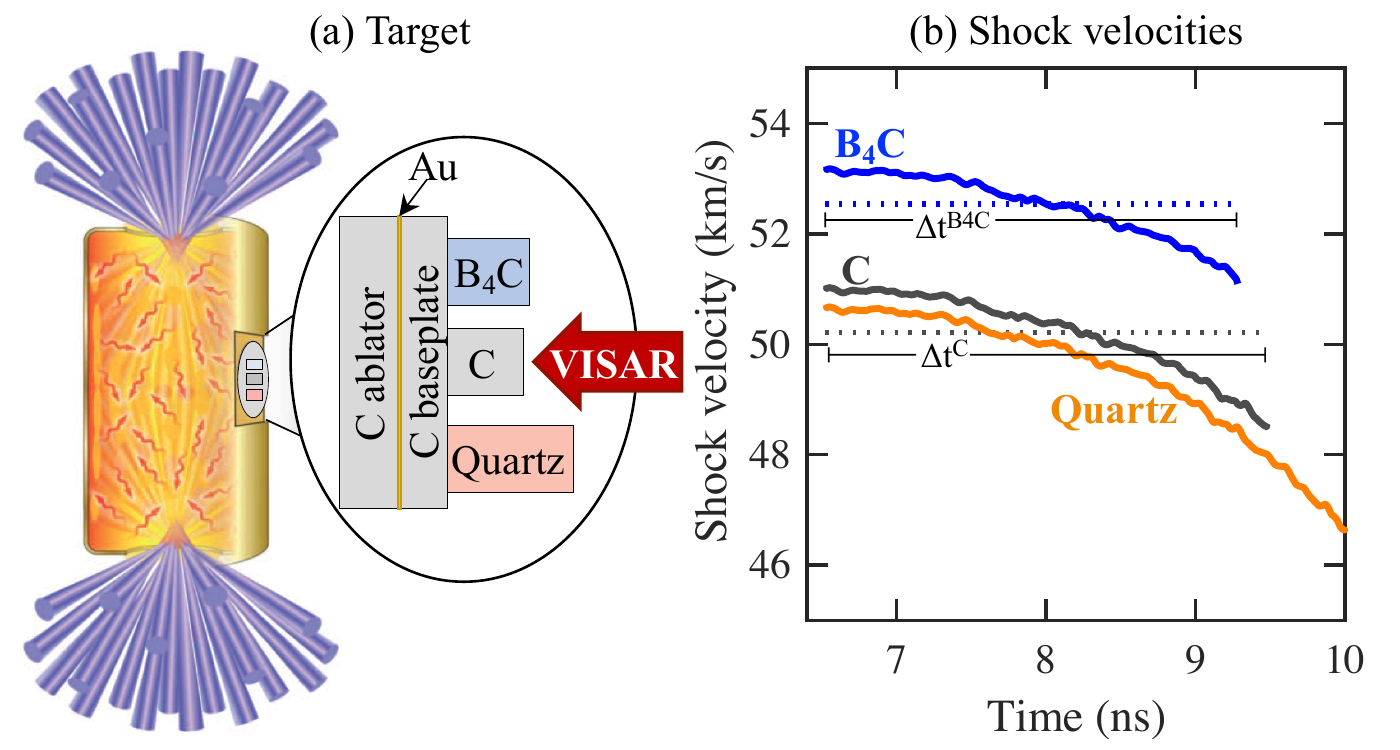}
	\caption{\label{fig:target} (a) Target design and (b) shock velocities in the boron carbide (B$_4$C), diamond (C), and quartz samples attached to the diamond baseplate for NIF shot N160414. Dotted lines in (b) show the average shock velocity in the samples determined from the measured thickness and shock transit time ($\Delta$t). Solid curves show the time-dependent shock velocity histories, measured using VISAR for quartz (orange) and determined using the nonsteady waves correction for B$_4$C (blue) and diamond (gray).}
\end{figure*}

\begin{table*}
	\caption{\label{table:expdata} B$_4$C Hugoniot data using the impedance-matching technique with a diamond standard. Shock velocities ($U_{\text{s}}$) at the diamond standard/sample interfaces were measured \textit{in situ} using VISAR for quartz ($Q$) and determined using the nonsteady waves correction for B$_4$C and diamond ($C$). $U_{\text{s}}^{\text{C}}$ and $U_{\text{s}}^{\text{B$_4$C}}$ were used in the impedance-matching analysis to determine the particle velocity ($u_{\text{p}}$), pressure ($P$), and density ($\rho$) on the B$_4$C Hugoniot. The average shock velocities ($\langle U_{\text{s}} \rangle$) determined from the measured thickness and shock transit times are also listed. The uncertainties for $\langle U_{\text{s}} \rangle$ are the same as those given for $U_{\text{s}}$.}
		\begin{ruledtabular} 
			\begin{tabular}{lcccccccc}
				Shot \# & $U_{\text{s}}^{\text{Q}}$ & $\langle U_{\text{s}}^{\text{C}}\rangle$ & $U_{\text{s}}^{\text{C}}$  & $\langle U_{\text{s}}^{\text{B$_4$C}}\rangle$ &$U_{\text{s}}^{\text{B$_4$C}}$ &$u_{\text{p}}^{\text{B$_4$C}}$& $P^{\text{B$_4$C}}$ & $\rho^{\text{B$_4$C}}$\\
				& (km/s) & (km/s) & (km/s) &  (km/s) & (km/s) & (km/s) & (Mbar) & (g/cm$^3$) \\
				\hline
				N160414 & 50.65 $\pm$ 0.25 & 50.22 & 51.00 $\pm$ 0.35 & 52.54 & 53.15 $\pm$ 0.46 & 36.86 $\pm$ 0.39 & 49.38 $\pm$ 0.59 & 8.22 $\pm$ 0.29 \\
				N161002 & 56.57 $\pm$ 0.25 & 55.35 & 56.43 $\pm$ 0.50 & 57.43 & 58.63 $\pm$ 0.43 & 41.60 $\pm$ 0.55 & 61.46 $\pm$ 0.86 & 8.68 $\pm$ 0.35 \\
				N170227 & 44.30 $\pm$ 0.25 & 44.61 & 45.39 $\pm$ 0.33 & 46.77 & 47.62 $\pm$ 0.30 & 31.98 $\pm$ 0.36 & 38.38 $\pm$ 0.46 & 7.67 $\pm$ 0.22 \\
				N170503 & 38.17 $\pm$ 0.25 & 38.62 & 39.18 $\pm$ 0.29 & 39.65 & 40.38 $\pm$ 0.25 & 26.84 $\pm$ 0.31 & 27.31 $\pm$ 0.33 & 7.52 $\pm$ 0.21 \\
				N170808 & 51.22 $\pm$ 0.25 & 50.38 & 51.13 $\pm$ 0.41 & 53.22 & 54.04 $\pm$ 0.36 & 36.82 $\pm$ 0.45 & 50.15 $\pm$ 0.65 & 7.91 $\pm$ 0.25 \\
				N180411 & 43.98 $\pm$ 0.25 & 44.49 &45.05 $\pm$ 0.37 & 46.56 & 47.15 $\pm$ 0.44 & 31.72 $\pm$ 0.40 & 37.67 $\pm$ 0.52 & 7.70 $\pm$ 0.28 \\
				N180611 & 39.67 $\pm$ 0.25 & 39.27 & 40.60 $\pm$ 0.40 & 40.68 & 42.15 $\pm$ 0.25 & 28.00 $\pm$ 0.43 & 29.74 $\pm$ 0.47 & 7.51 $\pm$ 0.26 \\
			\end{tabular}
		\end{ruledtabular}
\end{table*}

We present Hugoniot data for B$_4$C to 61 Mbar, exceeding the shock pressures achieved in previous experiments~\cite{Fratanduono_B4C} by a factor of eight. The new data were obtained from experiments at the NIF~\cite{NIF}, where the B$_4$C Hugoniot was measured relative to a diamond standard using the impedance-matching technique. The planar target package, which was affixed to the side of a laser-driven hohlraum, had a 200-\micron{}-thick diamond ablator, 5-\micron{}-thick gold preheat shield, a 100-\micron{}- or 125-\micron{}-thick diamond baseplate (the impedance-matching standard), and B$_4$C, diamond, and quartz samples as shown in Fig.~\ref{fig:target}(a). The surfaces opposite the drive of the diamond baseplate and smaller diamond sample were flash coated with 100 nm of aluminum to facilitate shock break out time measurements. Densities of the polycrystalline diamond, z-cut  $\alpha$-quartz, and B$_4$C were 3.515 g/cm$^3$, 2.65 g/cm$^3$, and 2.51 g/cm$^3$, respectively. The inner walls of the hohlraum were irradiated with 176 laser beams, which produced a $\sim$200 eV x-ray bath that drove a planar and nearly steady shock through the target package. The time-dependent shock velocity history in the quartz, measured using a line-imaging velocity interferometer for any reflector (VISAR)~\cite{Celliers_2004_VISAR}, showed only $~\pm$3\% variation from the average over the relevant time period of the experiment. The laser pulse duration, either 5 or 7.5 ns, and the total energy, between 519 and 820 kJ, varied shot-to-shot to produce high-pressure states in the B$_4$C spanning 27 to 61 Mbar. 

The shock velocities in the diamond baseplate (standard) and B$_4$C sample at the material interface are required to determine the pressure-density state on the B$_4$C Hugoniot using the impedance-matching technique. Average shock velocities through the smaller diamond and B$_4$C samples were calculated from their thicknesses, measured using a dual confocal microscope, and the shock transit times, measured using VISAR. The \textit{in situ} shock velocities in the B$_4$C and diamond samples were determined from the measured shock velocity history in the quartz using an analysis technique to correct for shock unsteadiness~\cite{Fratanduono_Nonsteady}. The average and \textit{in situ} shock velocities are shown in Fig.~\ref{fig:target}(b). The Hugoniot and release data for the diamond standard were determined using LEOS 9061, a multiphase EOS for carbon based on DFT-MD and PIMC calculations~\cite{Benedict_2014}. The experimental B$_4$C Hugoniot data are given in Table~\ref{table:expdata}. Further details on the experimental configuration and analysis techniques can be found in Ref.~\cite{Marshall_MMEOS}, which reports on quartz and molybdenum data that were acquired simultaneously with the B$_4$C data presented here.

\section{Results and Discussion}\label{sec:results}

\subsection{Hugoniot comparison}\label{subsec:b4chugcompare}
\begin{figure}
\centering\includegraphics[width=0.4\textwidth]{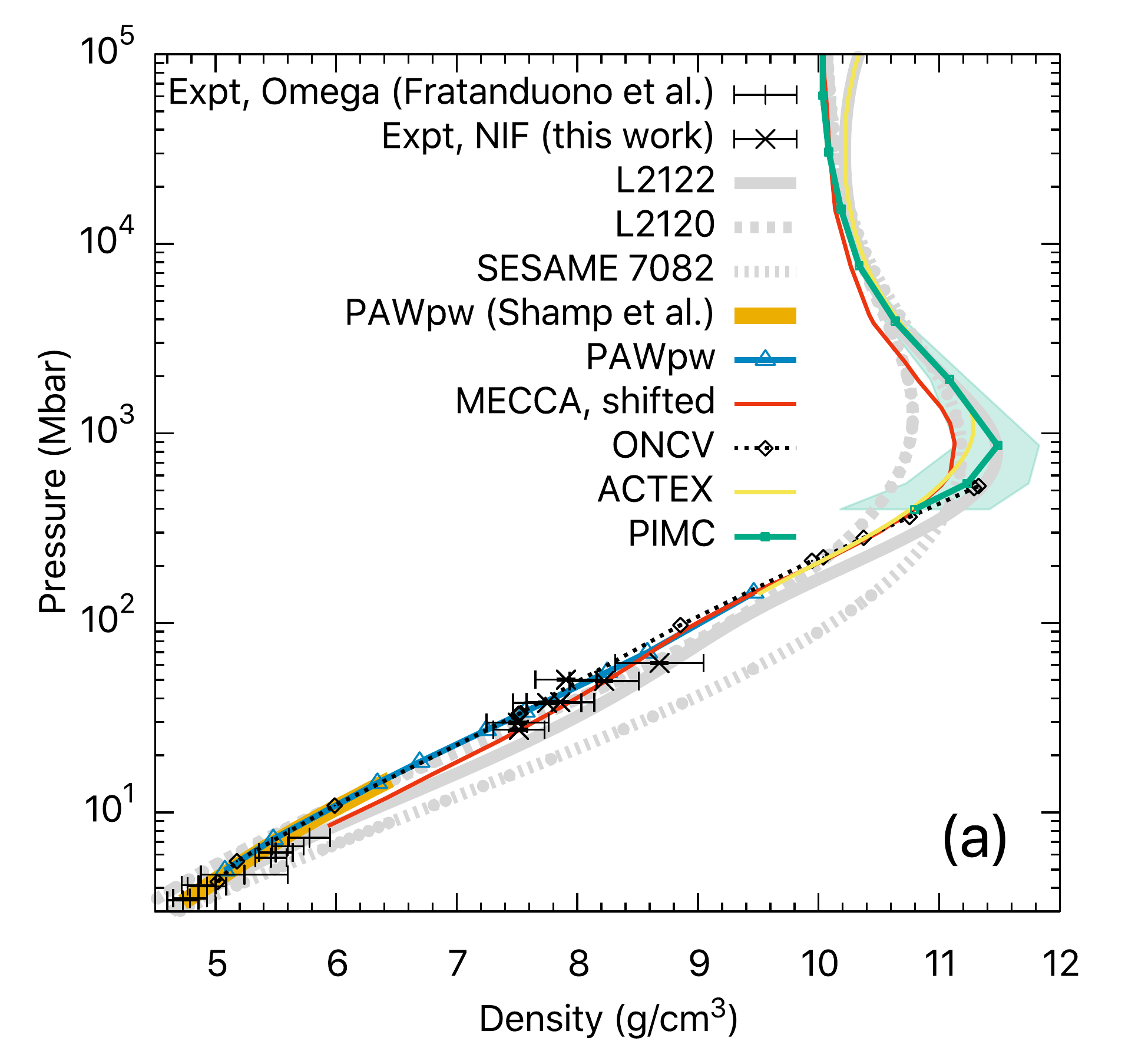}
\centering\includegraphics[width=0.4\textwidth]{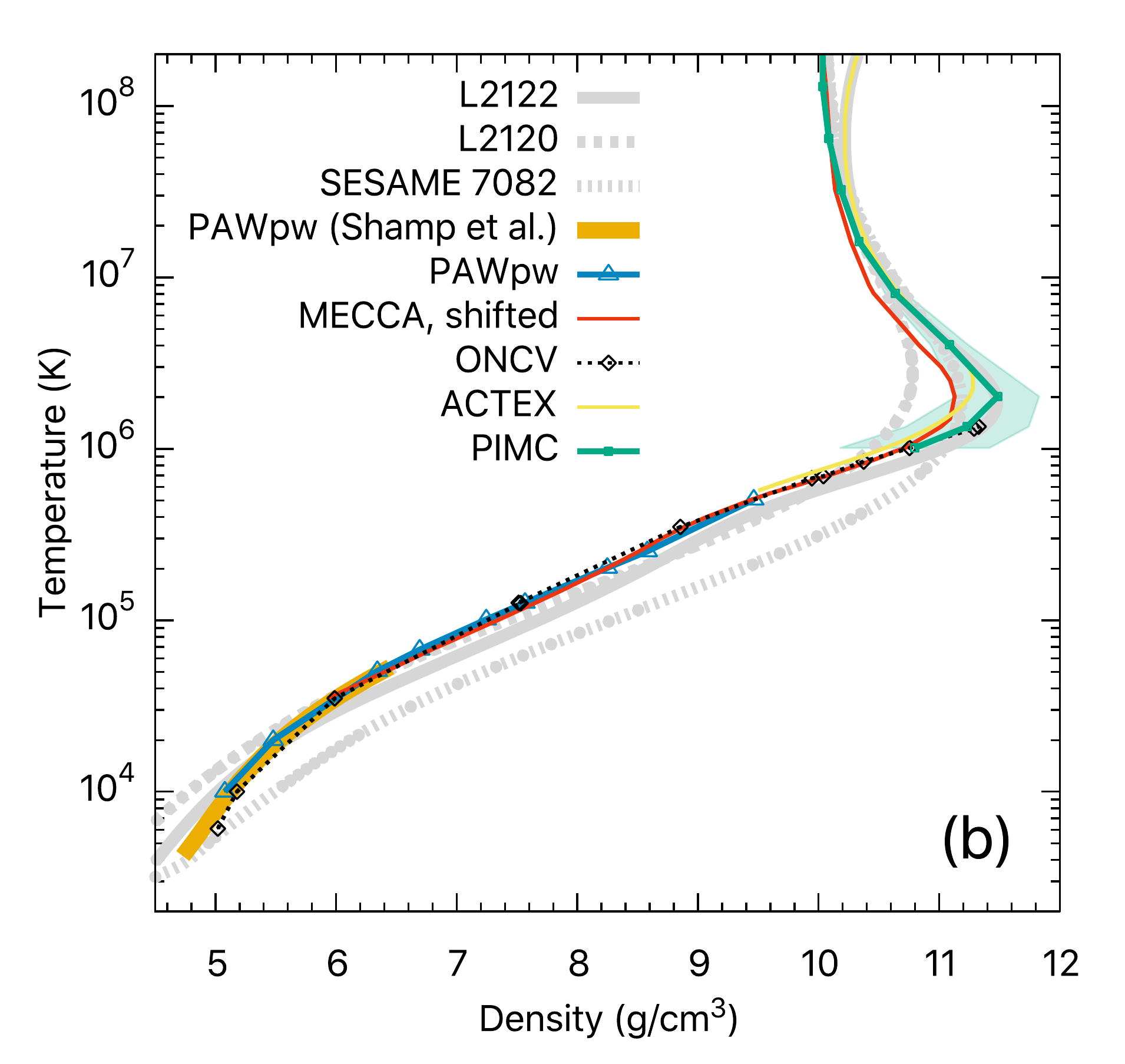}
\caption{\label{fig:prhohugexpt} Comparison of the Hugoniot of B$_4$C predicted by 
various simulations
and the LEOS/SESAME models in (a) pressure-density and (b) temperature-density representations. Also shown in (a) is our experimental data collected at the NIF and those by Fratanduono {\it et al.}~\cite{Fratanduono_B4C} at Omega laser facility. The shaded areas around the lower end of the PIMC curve represent 1$\sigma$ uncertainty in the corresponding Hugoniot density due to EOS errors. All pressures in our MECCA EOS table have been shifted up by 97.1 GPa, so that the value at ambient is zero. The deviation between PIMC/L2120 (and MECCA) and ACTEX/L2122 curves above $10^4$ Mbar is due to the electron relativistic effect, which is considered in ACTEX and L2122 but not in PIMC/L2120 (and not fully in MECCA). The initial sample density $\rho_i$=2.51 g/cm$^3$ for all the Hugoniot except that by Shamp {\it et al.}~\cite{Shamp2017}, which is 2.529 g/cm$^3$.}
\end{figure}

In this section, we compare our experimental measurements of the pressure-density Hugoniot of B$_4$C with our theoretical predictions. Figure~\ref{fig:prhohugexpt} compiles the experimental and theoretical Hugoniot curves in pressure-density
and temperature-density plots. 

The comparison in Fig.~\ref{fig:prhohugexpt} shows very good consistency 
between the measurements and the theoretical predictions. 
Assisted by the theoretical predictions, we estimate Hugoniot temperatures 
for the experimental data to be in the range of 1--5$\times$10$^5$ K. 
Our results also show that the PIMC and DFT-MD predicted Hugoniot are in
overall good consistency with LEOS 2122 (L2122).
Our calculations 
and the L2122 model predicts B$_4$C to have a maximum compression ratio 
of 4.55 at 9$\times10^2$~Mbar and 2$\times10^6$~K,
below which L2122 predicts B$_4$C to be slightly softer.
We also note that the pressure-density Hugoniots predicted by a different
Thomas-Fermi based tabular model L2120 is very similar at pressure ranges 
other than that around the compression maximum, at which the L2120 prediction is 
stiffer by $\sim$6\%. This can be attributed to the K-shell ionization 
that is fully captured by our calculations and the Purgatorio model L2122, 
while no atomic shell effects has been included in Thomas-Fermi models. 
In comparison, another Thomas-Fermi table (SESAME 7082), although 
reasonably agreeing with the low-pressure OMEGA data, can be clearly 
ruled out by our NIF data and computations. This may be attributed 
to the inaccuracies in the cold curve and the ion thermal model used 
in the table. The latest NIF gigabar (Gbar) experiments obtain Hugoniots of CH
and B near the compression maxima that agree with our PIMC 
calculations better than Thomas-Fermi predictions.~\footnote{A. Jenei, A. Kritcher, D. Swift, et al., {\it private communication}.}
We expect future, accurate experiments at Gbar
pressures to test our predictions for B$_4$C.

At 3--400 Mbar and 10$^4$--10$^6$ K, the Hugoniot curve obtained from 
MECCA and those from DFT-MD (PAWpw, ONCVpw, and FOE) agree remarkably 
well with each other. Because MECCA calculations are based on a static
lattice and the ion thermal contribution to the EOS is added following
an ideal gas model, the good consistency implies that the ion thermal 
contribution is dominated by the ion kinetic effect. 
We note that ACTEX predictions of the Hugoniot 
down to 6$\times$10$^5$ K and 140 Mbar
also agree very well with the DFT predictions.

The computational predictions are consistent with the NIF experimental 
data at pressures above 27 Mbar, as well as those
conducted at the Omega laser facility~\cite{Fratanduono_B4C} up to 5 Mbar.
However, at 5--10 Mbar, the experimental Hugoniot seems to be softer than 
DFT-MD predictions,
similar to findings by a previous DFT-MD study that was performed up to 15 Mbar~\cite{Shamp2017},
which might be attributed to chemical separation of the B$_4$C samples
as has been carefully explored for solid B$_4$C at low temperatures 
in Ref.~\onlinecite{Shamp2017}.

We note that, at temperatures of 1--4$\times10^6$ K, 
our PIMC data for B$_4$C have large errors (up to $\sim$2\%) 
because of the large computational cost and stochastic noise at these conditions.
The error quickly drops down 
with increasing temperature. We use a Monte Carlo approach to estimate the 
associated uncertainty in density along the Hugoniot by taking into 
account the errors in the PIMC data. The results are shown with the 
green shaded area in Fig.~\ref{fig:prhohugexpt}. 
It is clear that the PIMC Hugoniot is in excellent agreement with 
L2122 predictions and is consistent with those predicted by 
ACTEX and MECCA within the error bar. Slight differences of 
up to 2--3\% can be observed at 400--10,000 Mbar and 10$^6$--10$^7$ K. 
This may be due to the methodological difference between 
ACTEX/MECCA and PIMC/Purgatorio. 

\begin{figure}
\centering\includegraphics[width=0.4\textwidth]{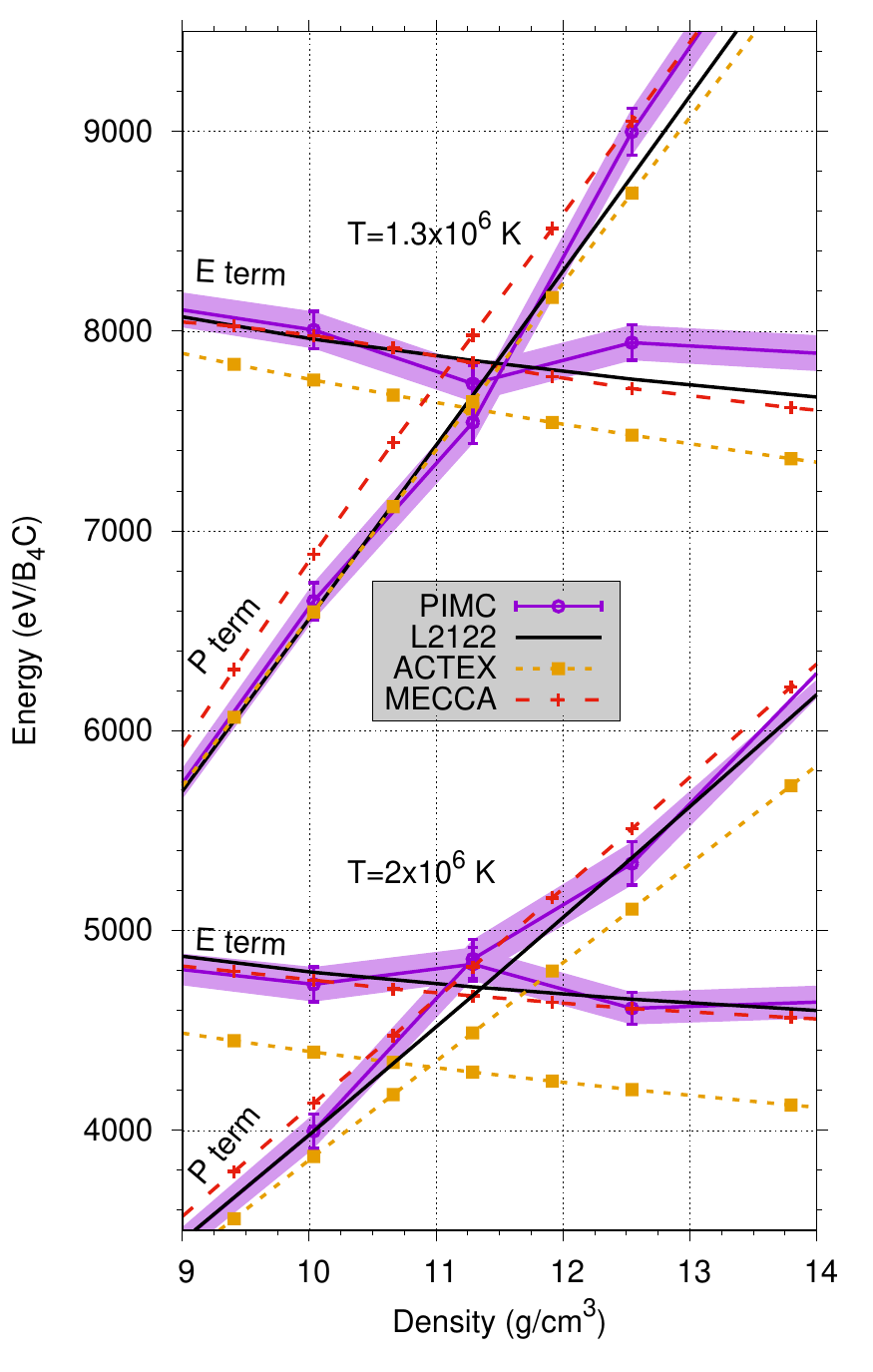}
\caption{\label{fig:prhohugtheory} Comparison of the energy and pressure terms of the Hugoniot function for B$_4$C from different theories and LEOS models at two temperatures around the compression maximum. The shaded area denote the standard error of the PIMC EOS. }
\end{figure}

In order to better understand the origin of the differences 
at the compression maximum, we compare
in Fig.~\ref{fig:prhohugtheory}
the energy term $E-E_i$ and the pressure term $(P+P_i)(V_i-V)/2.0$,
where $(E, P, V)$ and $(E_i, P_i, V_i)$ respectively denote the internal energy, pressure, and volume
of B$_4$C under shock and in its initial state (300 K and 2.51 g/cm$^3$),
of the Hugoniot function from PIMC, ACTEX, MECCA, and LEOS 2122 along
two isotherms 1.3$\times10^6$ K and 2$\times10^6$ K.
The cross point between the curve of the energy term and 
that of the pressure term 
gives the Hugoniot density at the corresponding temperature.
Our comparison shows that the internal energy slowly decreases
while the pressure term dramatically increases, 
as the density increases from 9 to 14 g/cm$^3$.

Due to the high computational expense of PIMC simulations at low-temperature conditions, our PIMC data at low temperatures exhibit significantly larger error bars and stochastic noise than the higher temperature results.
The error bars of
the PIMC data lead to estimations of the $1\sigma$ uncertainty in 
Hugoniot density, as is shown with shaded green areas
in Fig.~\ref{fig:prhohugexpt}.
L2122 and PIMC agree well with each other
in both energy and pressure, explaining the excellent consistency
between their predicted Hugoniots.
MECCA pressures are slightly higher than PIMC, whereas energies are
similar, therefore the Hugoniot density is also lower.
In comparison to PIMC, ACTEX energies are lower, 
while pressures are similar at 1.3$\times10^6$ K
but lower at 2$\times10^6$ K,
therefore the Hugoniot densities from ACTEX are also lower.

\subsection{EOS comparison}\label{subsec:b4ceoscompare}

\begin{figure}
\centering\includegraphics[width=0.4\textwidth]{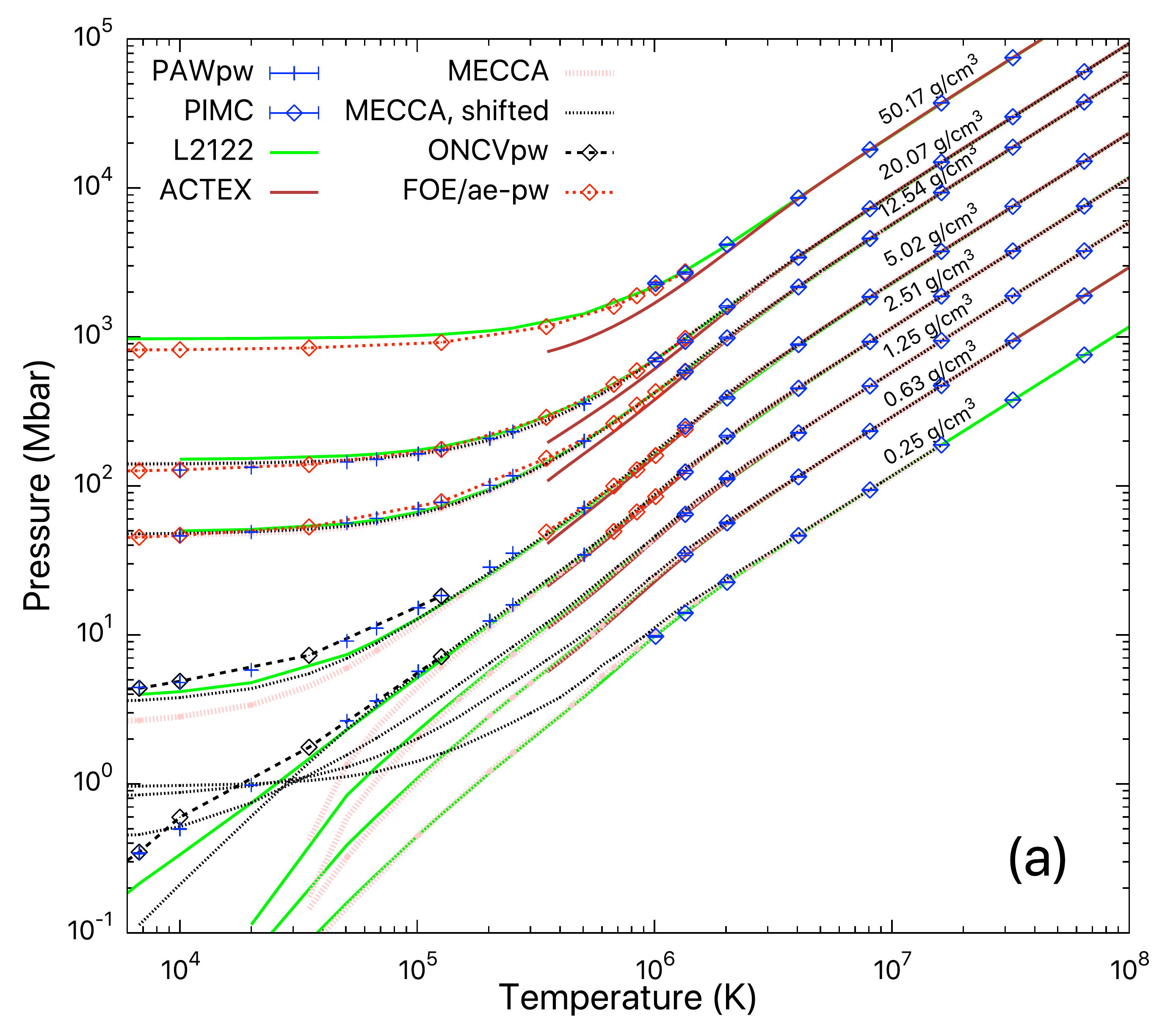}
\centering\includegraphics[width=0.4\textwidth]{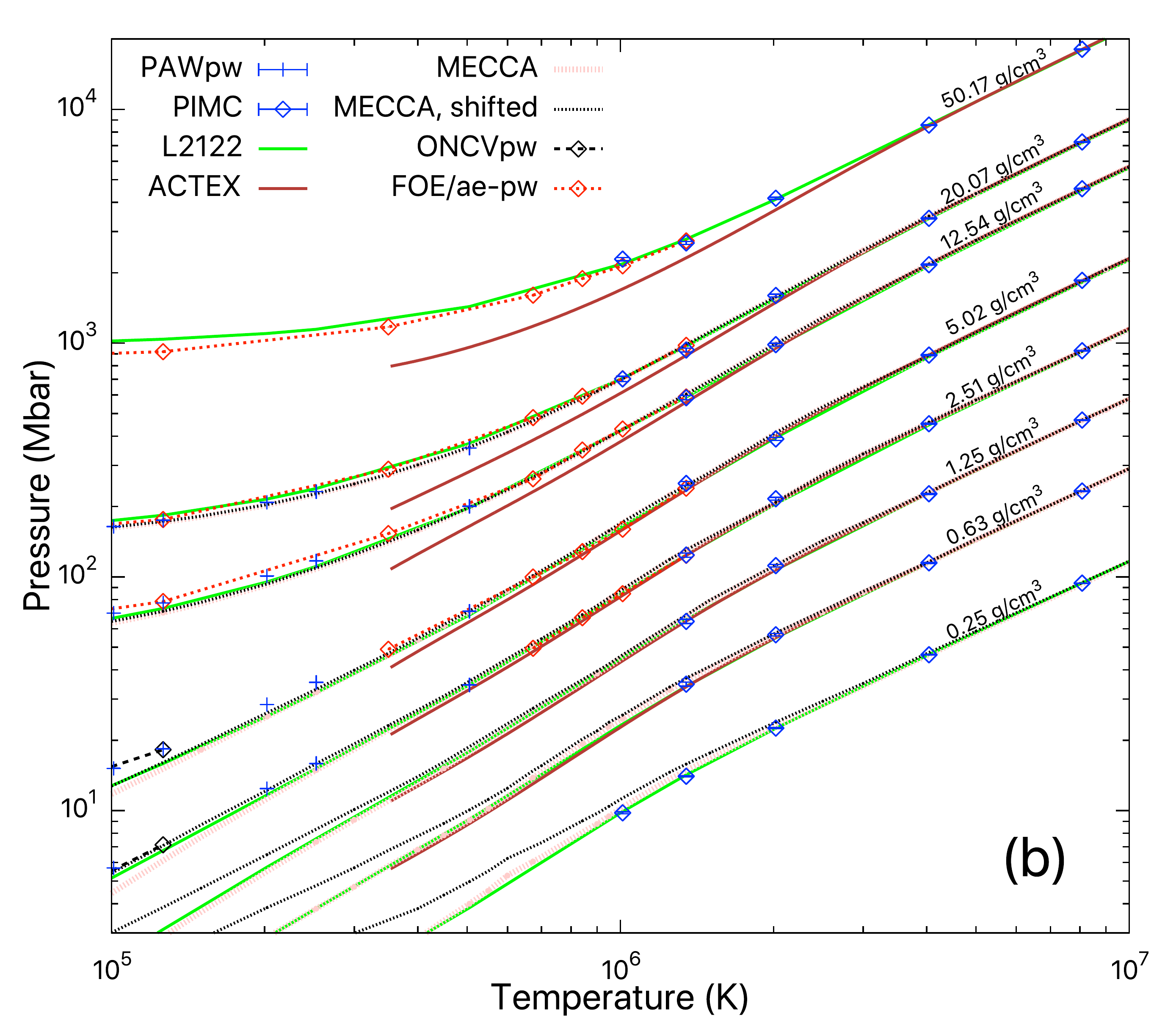}
\caption{\label{fig:pisochore2} Comparison of the pressure-temperature profiles of B$_4$C along several isochores from PIMC, DFT-MD [PAWpw, frozen 1s; ONCVpw, frozen 1s; FOE or pw, all-electron(ae)], ACTEX, MECCA, and L2122. Also included is a set of MECCA data that have been shifted up by 97.1 GPa, so that the value at ambient is zero. Subplot (b) is a zoom-in version of (a).}
\end{figure}

The principle Hugoniot samples a specific pathway in the phase space
from 2.5 to 11.5 g/cm$^3$ accompanied by increasing temperatures.
These conditions are very important because the corresponding states 
are reachable using shock experiments.
However, off-Hugoniot states, as those simulated in the present work,
also play vital role in hydrodynamic simulations and the underlying physics
can be different. We therefore make detailed comparisons of the EOS
among various methods in this subsection.

The pressure-temperature data along several isochores from our
calculations are compared in Fig.~\ref{fig:pisochore2}.
At 4$\times10^6$ K and above, all our methods
(PIMC, ACTEX, and MECCA) agree and are consistent with
the L2122 model.
This is understandable because the system is approaching
the limit of a fully-ionized classical plasma,
which is accurately described by PIMC, ACTEX, and the DFT
methods MECCA and Purgatorio.

At lower temperatures, the different ways of employing
DFT-MD (PAWpw, ONCVpw, and FOE) give the same EOS and 
consistent trend with the PIMC data. Several differences
are noteworthy when other methods (ACTEX, MECCA, and L2122) 
are considered:
(1) ACTEX pressures being lower than others, 
more so at higher densities;
(2) MECCA pressures being significantly different from 
L2122 at 5 g/cm$^3$ and below, in particular at $T<10^5$ K;
(3) with a rigid shift-up of 97.1 GPa (so that the ambient pressure is zero), MECCA pressures agree better with L2122 at ambient density and above, 
but worse at lower-than-ambient densities;
and (4) FOE pressures gets slightly lower than L2122
for densities higher than 25 g/cm$^3$.

\begin{figure*}
\centering\includegraphics[width=0.6\textwidth]{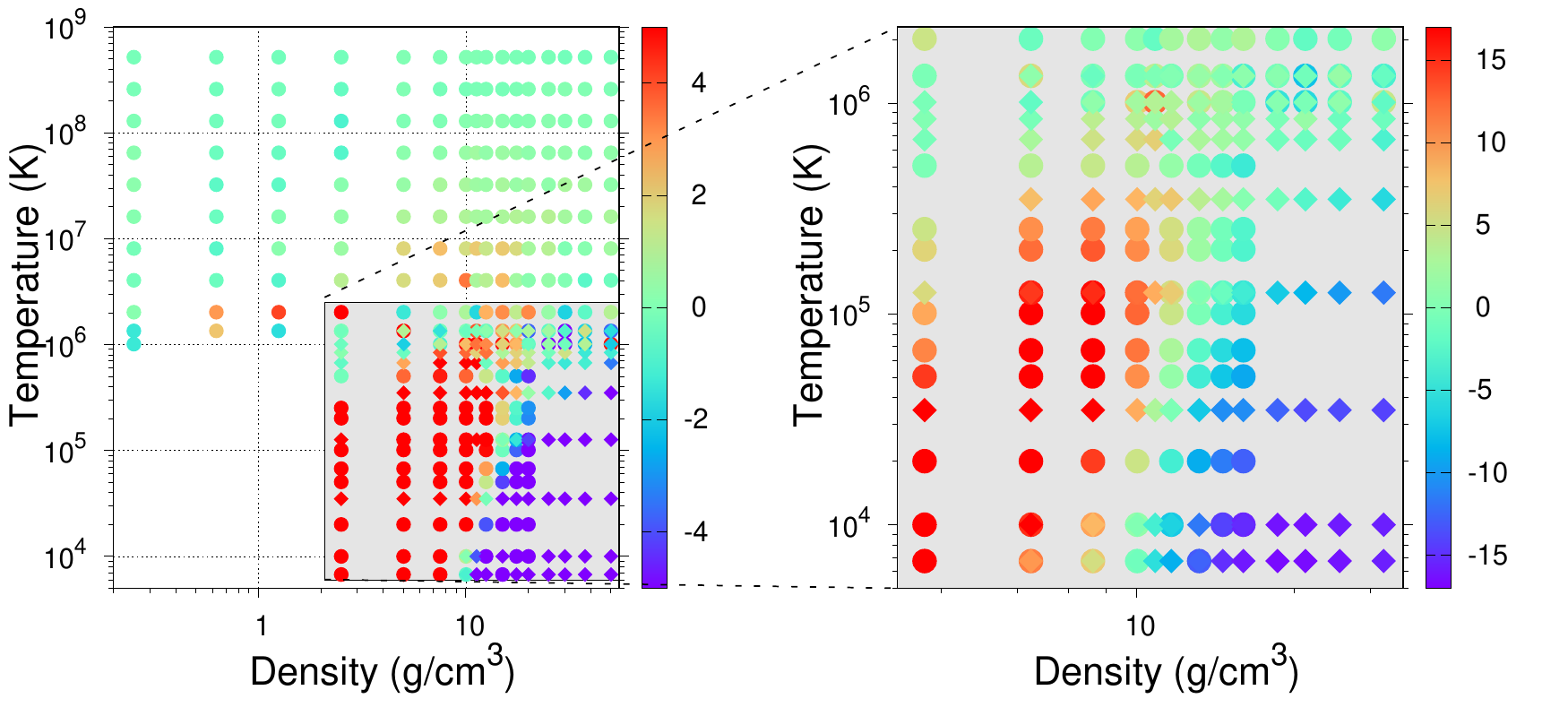}
\caption{\label{fig:pdiff} Percent difference in pressure of B$_4$C between PIMC/PAWpw (in spheres) or ONCV (in diamonds) and L2122.}
\end{figure*}

Figure~\ref{fig:pdiff} focuses on the differences 
between the first-principles PIMC/DFT-MD data and L2122
$\Delta P=(P^\text{FP}-P^\text{L2122})/P^\text{L2122}*100\%$. 
The agreement is well within 3\% for all densities studied
presently and temperatures above 4$\times10^6$ K.
At lower temperatures, $\Delta P$ varies between $\pm17\%$
depending on the density---DFT-MD pressures are
in general higher at densities below 10 g/cm$^3$
and lower above.
$|\Delta P|$ becomes smaller than 10\% and gradually vanishes
when temperature increases to 3.5$\times10^5$ K or above.
PAWpw and ONCVpw/FOE predictions are overall the same.
FOE smoothly bridges with PIMC predictions at 10$^6$~K.

\begin{figure}
\centering\includegraphics[width=0.4\textwidth]{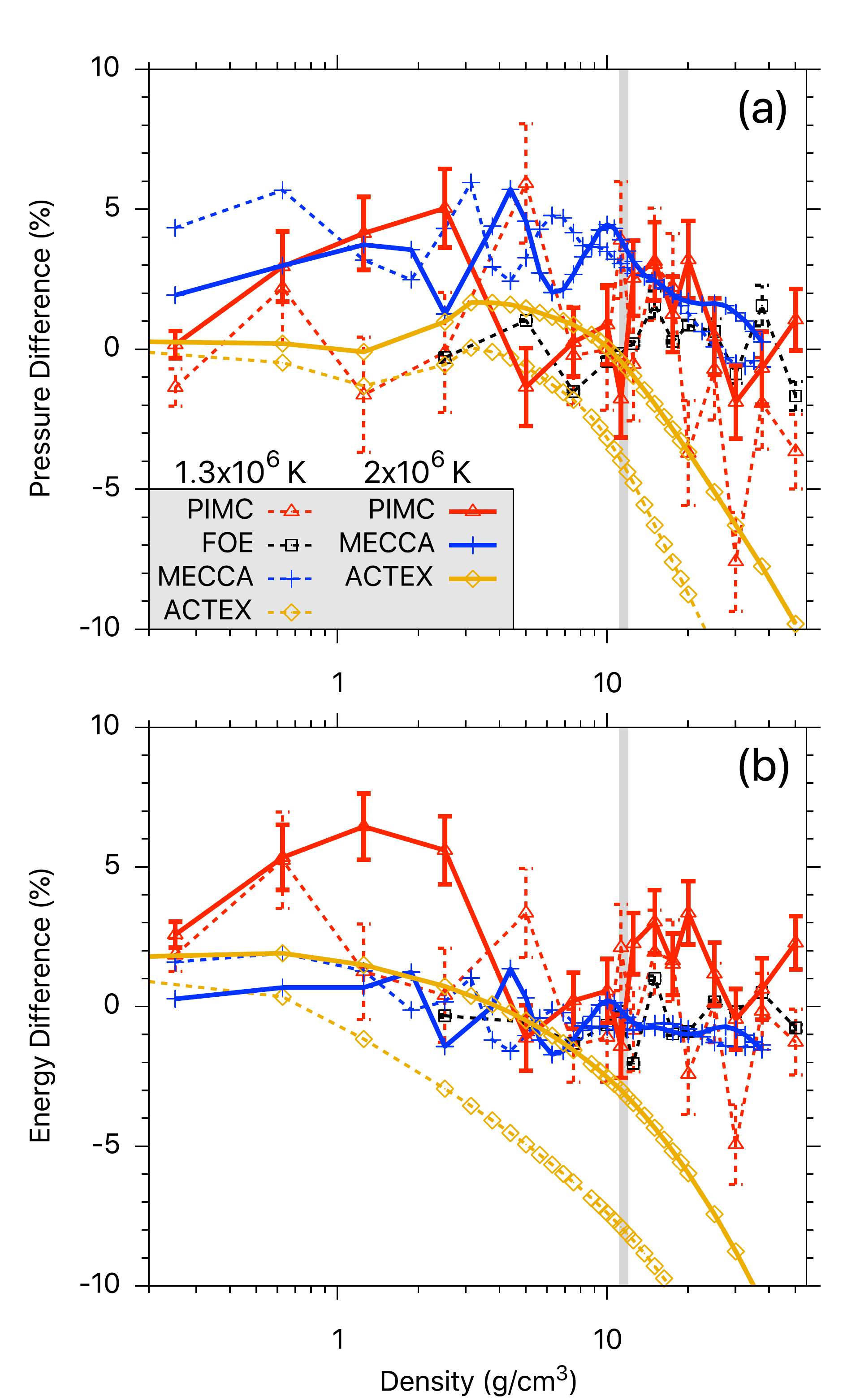}
\caption{\label{fig:allvsl2122} EOS differences of PIMC (red), FOE (black), MECCA (blue), and ACTEX (yellow) relative to LEOS 2122 along two isotherms [1.3$\times10^6$ (dashed curves) and 2.0$\times10^6$ K (solid curves)]. Because of the different references chosen in the EOS datasets, all energies have been shifted by the corresponding values at ambient condition (2.5087~g/cm$^3$ and 300~K). The pressure differences are normalized by the corresponding LEOS 2122 values; the energy differences are normalized by the fully-ionized ideal gas values ($46.5k_BT$ per B$_4$C).
The statistical error bars correspond to the 1$\sigma$ uncertainty of the FOE and PIMC data. The gray vertical bar at 11.54 g/cm$^3$ denotes the maximum Hugoniot density according to LEOS 2122 and PIMC.}
\end{figure}

We also compare the pressures and energies from our
different computations with those from L2122.
The results along two isotherms 1.3$\times10^6$ and 2$\times10^6$ K
are shown in Fig.\ref{fig:allvsl2122}.
We find that PIMC, MECCA, and FOE agree with each other 
to within 5\%, which is comparable to what we found about
differences between PIMC and DFT-MD in previous work on 
B~\cite{Zhang2018b}, BN~\cite{Zhang2019bn}, and hydrocarbon 
systems~\cite{Zhang2018,Zhang2017b}.
The cross validation of the different DFT methods and their
consistency with PIMC predictions confirm that both the PIMC 
and the DFT-MD approaches, albeit carrying approximations in 
each, are reliable for studying the EOS of warm dense matter.
Our ACTEX data also show remarkable consistency 
(e.g., $<2\%$ at 2$\times10^6$ K) with L2122 at
densities below 10 g/cm$^3$.
However, the ACTEX data get way too low at higher densities,
which is due to breakdown of the ACTEX method when the 
two-body term at order 2 in the activity becomes comparable to the Saha term,
similar to what has been found for BN~\cite{Zhang2019bn}.

\subsection{Modifications to L2122 and 1D hydrodynamic simulations}\label{sec:hydro}

\begin{figure}
\centering\includegraphics[width=0.4\textwidth]{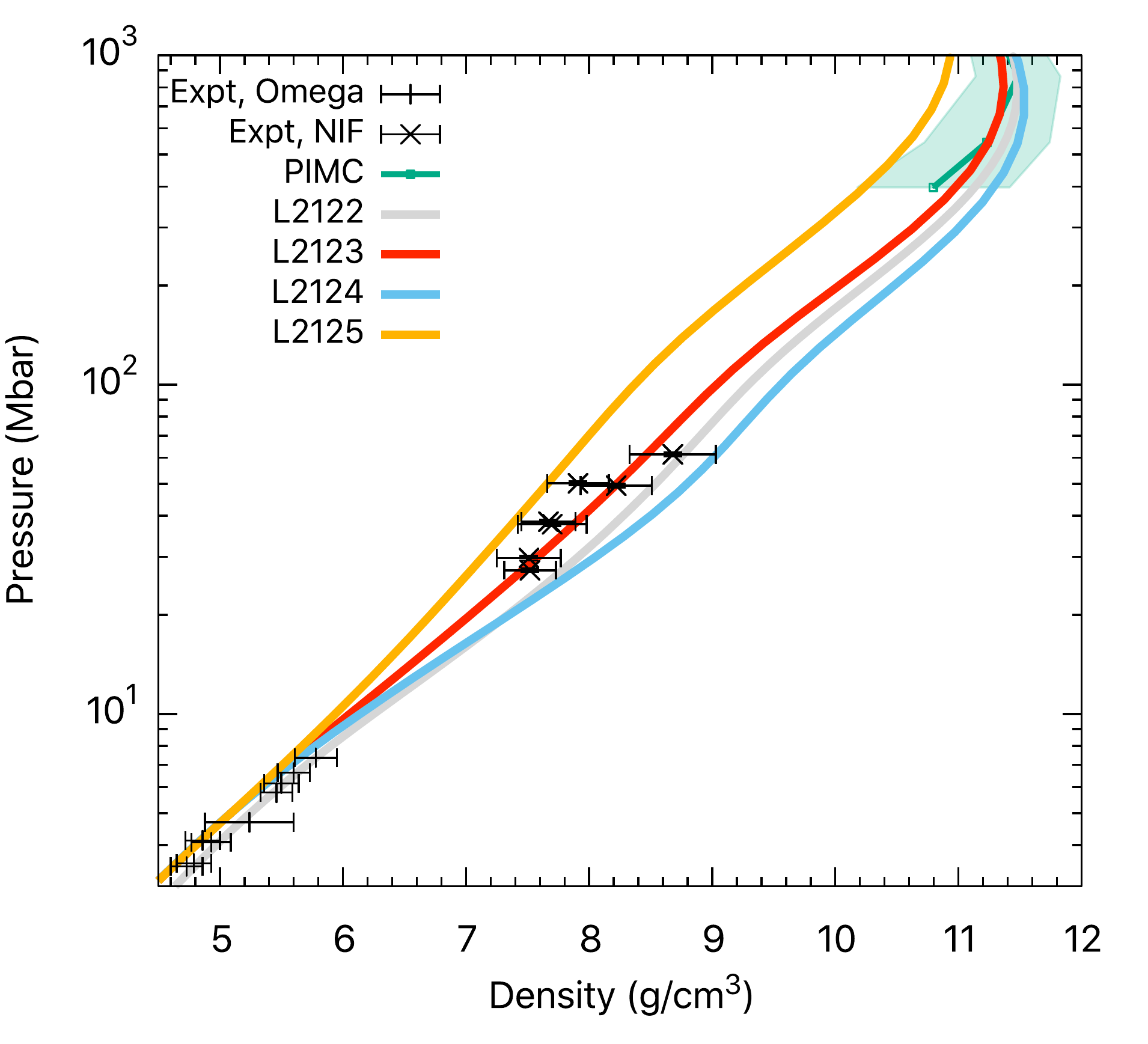}
\caption{\label{fig:newmods} {Comparison of the Hugoniot ($\rho_i$=2.51 g/cm$^3$) of B$_4$C from newly constructed QEOS models (L2123, L2124, and L2125) and those from experiments, PIMC simulations, and L2122.}}
\end{figure}

We have shown in Fig.~\ref{fig:prhohugexpt} that L2122 predicts slightly softer behavior for B$_4$C at 5--500 Mbar, despite the overall good consistency, in comparison with our first-principles and experimental Hugoniot.
We have thus created three new models for the B$_4$C EOS, with the intent to span the range of Hugoniot behavior that is in better agreement with the experimental data from both NIF and Omega.
Recent advances in ICF design methodologies that leverage Bayesian inference techniques to find most probable physics models based on a range of
experimental outcomes\cite{Gaffney2019} and recent interest in B$_4$C as an ablator for such experiments motivated us to create this range of possible EOS models rather than just a single table.
By considering the range of reasonable EOS models for B$_4$C
as obtained from our above comparisons of theoretical methods
and experimental uncertainty, we developed these three new tables by making modifications to the
Gr\"{u}neisen parameter within the QEOS methodology.  

The Hugoniot curves corresponding to the new models (L2123, L2124, and L2125)
are shown in Figure~\ref{fig:newmods}, along with the experimental data.  The PIMC Hugoniot with error bars is also shown.  The new baseline model (L2123)
has a slight modification to the Gr\"{u}neisen parameter, which determines the ion thermal EOS, to bring it into better agreement with both sets of experimental data.
L2124 and L2125 have modified forms of the Gr\"{u}neisen parameter that span the range of the experimental error bars.  Both L2123 and L2124 (the softer model) closely track
L2122 near peak compression, whereas the L2125 (the stiffer model) shows significantly modified behavior near peak compression.

We applied these new models to 1D hydrodynamic simulations of a polar direct drive fusion experiment based on previous studies.\cite{Ellison_2018,whitley2020comparison}
For this study, we kept the capsule diameter constant at 3000 $\mu$m and set the gas pressure to 8 atm of D$_2$ at room temperature.
We used a flux limiter=0.0398 and a square pulse shape with peak power set to 280~TW. The pulse duration was chosen such that
476~kJ of energy would be available from the laser.  Due to geometric losses, we assumed that the maximum absorption of energy would correspond to 75~\% of the total energy available.
Similar to our previous work on boron~\cite{Zhang2018b}, we found that the EOS variations we considered here did not produce significant differences in the fuel areal density, peak ion temperature,
or ablator areal density in these direct drive simulations.  

In order to expand this sensitivity study to situations that might be more relevant to future neutron source development studies,\cite{Yeamans_2020}
we also examined the neutron yield vs. ablator thickness for each of the three EOS models.  
Interestingly, all four EOS models (L2122--L2125) predict similar profiles for the neutron yield with
a peak yield that occurs
around an ablator thickness of 7.5~$\mu$m. Differences between the models are all within 1\% for thin ($<$10~$\mu$m) ablators. For ablator thickness between 10--25 $\mu$m, we found the neutron yield from L2123 remains similar ($<$0.2\%), while that from L2124 and L2125 deviate by up to 3\%, in comparison to L2122. These results demonstrate the availability of these models for use in future studies in ICF design with novel ablators.

\section{Discussion}\label{sec:discuss}

For the sake of benefiting future EOS development, 
high energy density physics, and warm dense matter studies, 
we hereafter discuss the physical origins of the EOS differences
shown above from electronic-structure and QEOS points of view.

\subsection{Finite size effects}\label{subsec:finitesizeeffect}

\begin{figure}
\centering\includegraphics[width=0.5\textwidth]{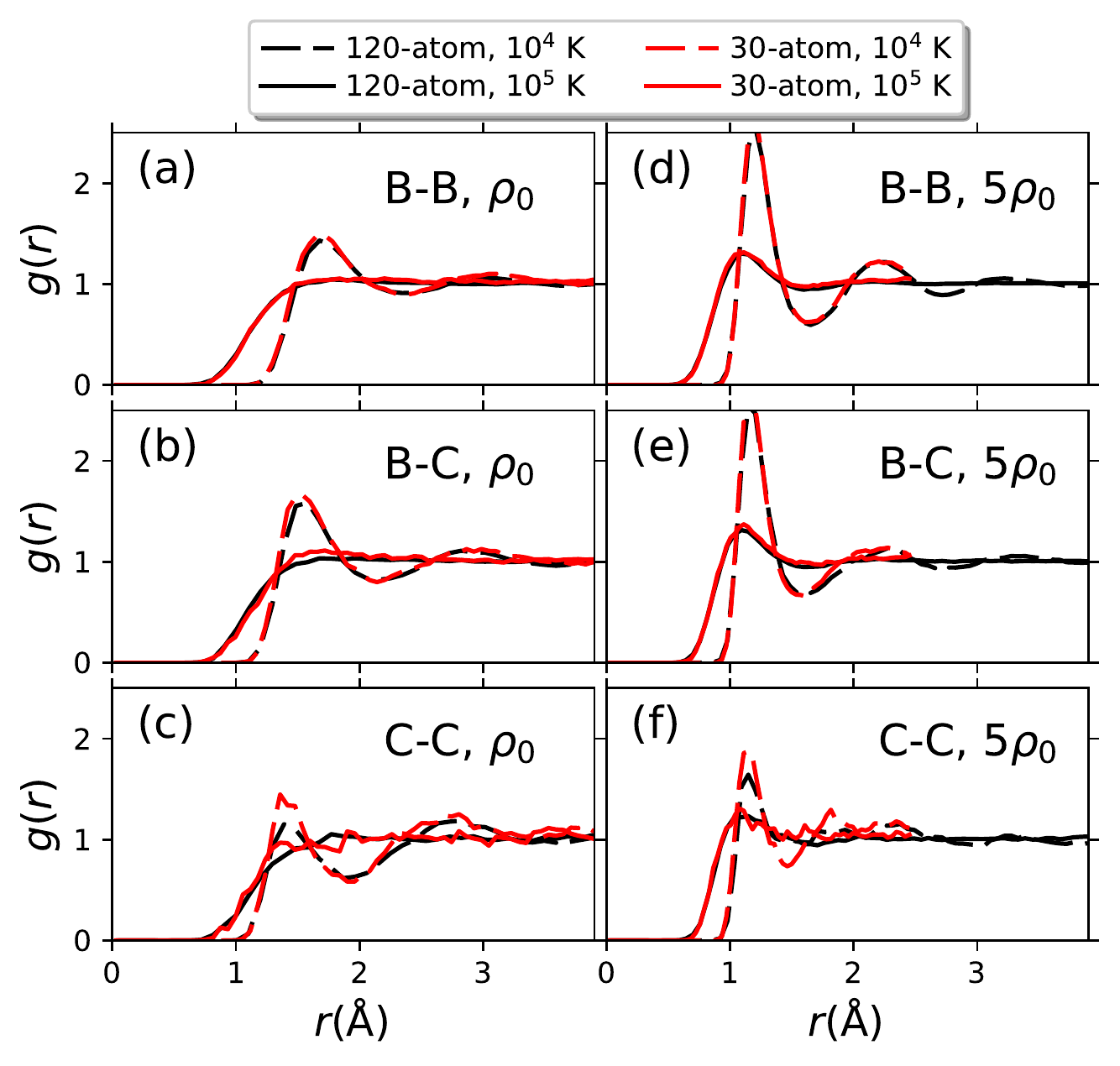}
\caption{\label{fig:gr}  Comparison of the nuclear pair correlation function obtained from DFT-MD (PAWpw) for B$_4$C using 30-atom (red) and 120-atom (dark) cells at two different densities and two temperatures. The reference density $\rho_0$ is 2.5087 g/cm$^3$. }
\end{figure}

Our first-principles calculations PIMC, PAWpw, ONCVpw, and FOE
implement the standard way of simulating liquids~\cite{Allen1987}, 
which considers a finite number of atoms in a cubic box and under 
the periodic boundary condition. The finite-cell size effects
have been carefully addressed in our DFT-MD simulations by
choosing large-enough cells with 120 atoms for all temperatures 
up to 2.5$\times10^5$ K ($\sim20$ eV). This is much higher than
the chemical bonding (typically about a few eV) is allowed,
which justifies the usage of 30-atom cells for all simulations at higher temperatures. In order to show this, Fig.~\ref{fig:gr} compares
the nuclear pair correlation function at two different temperatures
(10$^4$ and 10$^5$ K) and two different densities (2.5 and 12.5 g/cm$^3$)
using two different cells sizes (30 and 120 atoms), from our
PAWpw calculations. The results show remarkably good agreement
in the features of $g(r)$ using 30-atom cells with those using
the much larger 120-atom cells even at the relatively 
low-temperature ($10^4$ K), high-density (12.5 g/cm$^3$) condition. 
This is different from our recent findings for BN,
which show stronger size dependence at similar conditions,
and is probably due to larger polarization effects in BN
than in B$_4$C.
Moreover, structures can be clearly seen in the pair correlation plot
at 10$^4$ K, which are signatures of chemical bonding.
At $10^5$ K, these structures smooth out and the $g(r)$ becomes more 
ideal-gas like, which validates the ideal mixing 
approximation in multi-component average-atom EOS
approaches~\cite{Zhang2017b,Zhang2018}.

\begin{figure}
\centering\includegraphics[width=0.4\textwidth]{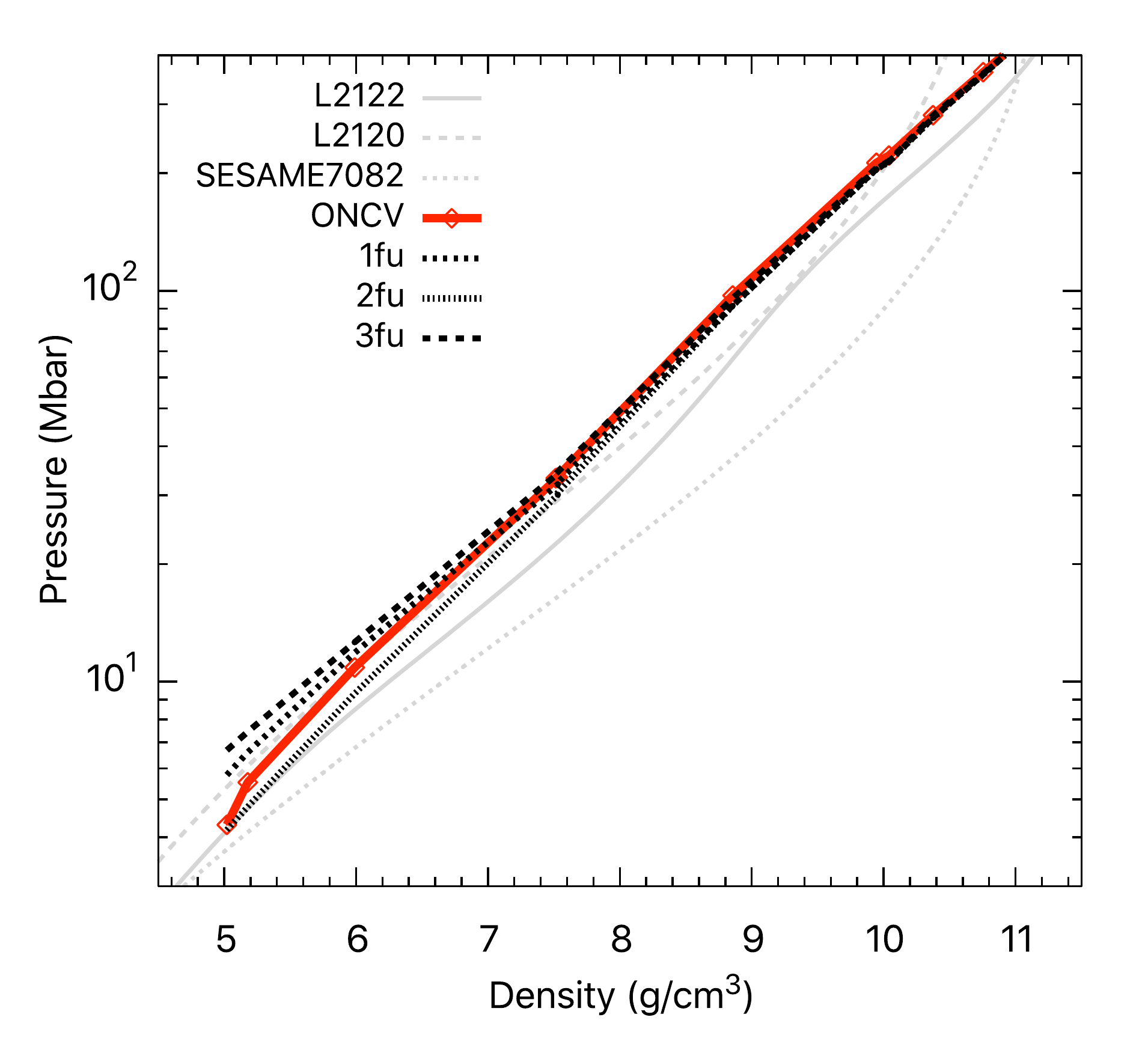}
\caption{\label{fig:prhohugfs} Comparison of pressures 
from single-snapshot calculations using various cells [1 formula unit (fu): 5-atom cell; 2 fu: 10-atom cell; 3 fu: 15-atom cell] at the ONCV Hugoniot densities and temperatures.
The Hugoniot from three EOS models are also shown for comparison.}
\end{figure}

At temperatures below $10^5$ K, chemical bonding has to be described
using reasonably big simulation cells so that the EOS can be accurately
obtained. In Sec.~\ref{subsec:b4ceoscompare}, we show that
MECCA calculations using a 5-atom cell produce a pressure (-97.1 GPa) that is significantly
different from 1 bar at ambient condition, and therefore
a rigid shift in pressure
for the MECCA EOS table has to be applied to improve the agreement
between MECCA and DFT-MD Hugoniots. 
It is worthwhile to investigate the effect of using such small sizes in more depth by making comparisons with slightly larger ones.

We constructed three structures consisting of 5, 10, and 15 atoms respectively~\footnote{The 5-atom structure has four B forming a face-centered cubic lattice and one C at the body center of the cell; the 10-atom structure consists of two C forming a body-centered cubic lattice that interconnects with 8 B that occupy the $(\pm1/4,\pm1/4,\pm1/4)$ and $(\pm3/4,\pm3/4,\pm3/4)$ sites and form a simple-cubic lattice; the 15-atom structure is a pseudo-rhombohedral phase of B$_4$C consisting of a boron icosahedra and a carbon chain.}, 
and performed 
additional pw-based single-snapshot calculations using all-electron ONCV potentials
along the density-temperature Hugoniot predicted using the ONCVpw/FOE EOS.
The pressure data as a function of density from the new ONCV calculations
are compared in Fig.~\ref{fig:prhohugfs}).
The results show that using 10-atom cells brings the pressure down relative to that
using 5-atom cells. However, 
using larger,
15-atom cells leads pressure to the opposite direction, instead of
approaching the converged values. The differences as signatures of
ion thermal and cold-curve effects on the EOS of B$_4$C are observable
along the Hugoniot at densities up to 9 g/cm$^3$, 
which is $\sim$100 Mbar and $\sim3\times10^5$ K.

\subsection{Roles of kinetic and interaction effects from ions and electrons}\label{subsec:kinvsinteffects}

\begin{figure}
\centering\includegraphics[width=0.42\textwidth]{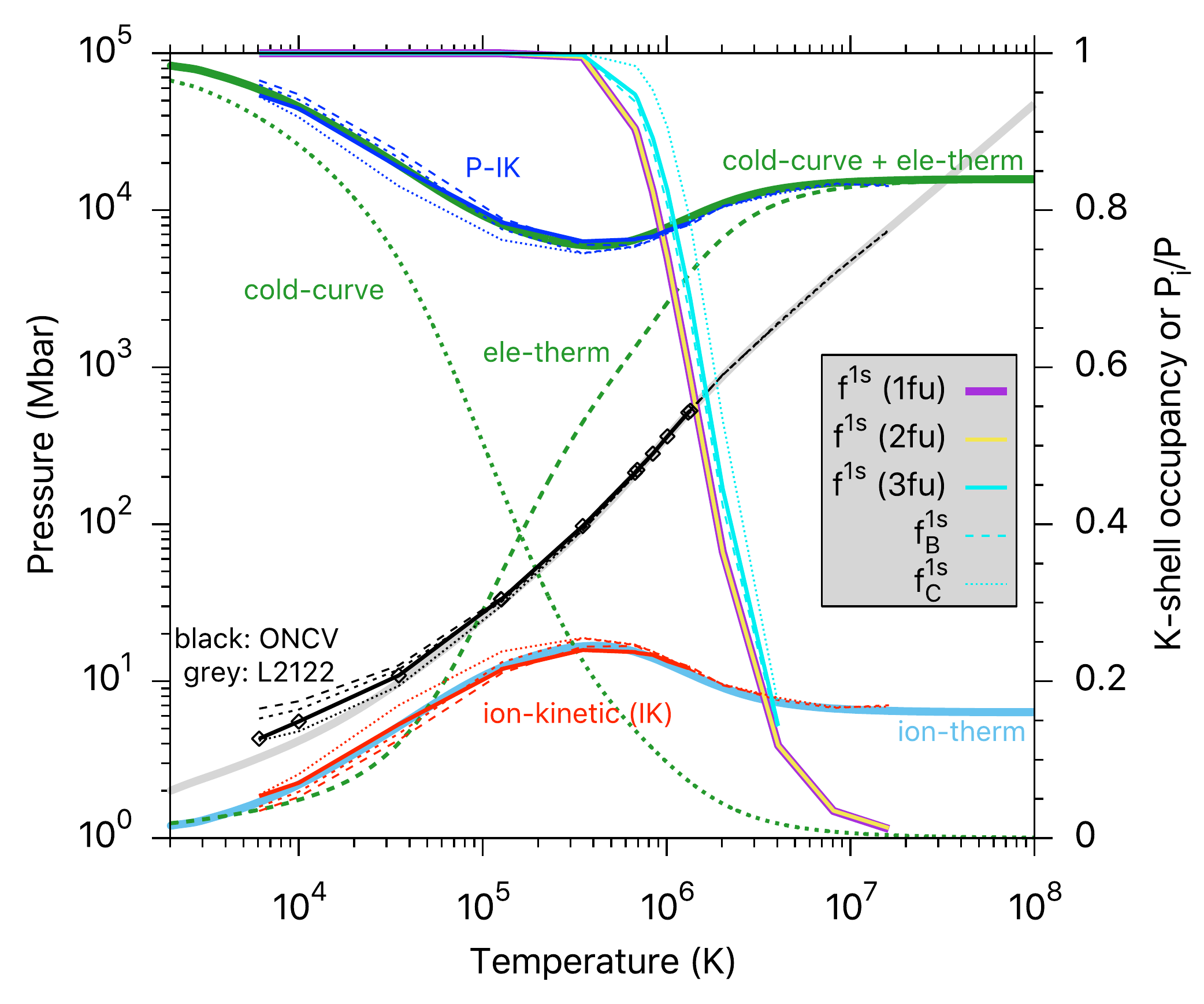}
\caption{\label{fig:bnpcomp} Fractional decomposition of pressure (right axes) in the LEOS 2122 model and ONCV calculations along their respective Hugoniots (black/grey curves, left axes). Also shown are the K-shell occupancy (right axes) of boron and carbon atoms and the average values as obtained from ONCV calculations using different cell sizes (shown in the legend). The ONCV pressures calculated at the Hugoniot temperature and density conditions and using smaller cells [short dashed: 1 formula unit (fu) (5-atom cell); dotted: 2 fu (10-atom cell); dashed: 3 fu (15-atom cell)] are shown for comparison.}
\end{figure}

In order to clarify the roles of kinetic and interaction effects
and those from the ions and from the electrons, we performed
additional analysis of our pw-based all-electron ONCV calculations.
The calculations allow decomposing the total pressure
into an ion-kinetic (IK)  term, which is calculated using the 
ideal gas model, and a remaining term (P-IK)(Fig.~\ref{fig:bnpcomp}). 
In comparison to
the QEOS way of decomposing the L2122 Hugoniot pressure into 
ion-thermal, electron-thermal, and cold curve components,
we find that the IK contribution is overlapping with the ion-thermal term in L2122.

In addition, we find that the temperatures at which finite cell size effects
are significant, as characterized by the differences between
solid and dashed curves,
overlap with those at which the cold-curve
surpasses the ion-thermal contributions.
The turn-over point $T_\text{t}$, $\sim3\times10^5$ K for B$_4$C, may 
be interpreted as a conservative estimation of the uppermost
temperature at which finite-size effect remains significant in a
theoretical computation, or the lowermost temperature
at which an average-atom approach is feasible.
Below $T_\text{t}$, interactions are so significant that
the ideal mixing approximation becomes less reliable
and a large simulation cell is required for the accuracy of
computations.

As temperature increases to a critical value $T_\text{c}$
where K-shell ionization starts, the electron-thermal
contribution becomes dominant. This leads to
a saddle point in the IK and the P-IK curves in Fig.~\ref{fig:bnpcomp}.
Our present calculations show $T_\text{c}=3\times10^5$ K
for B$_4$C, which is close to what we previously obtained 
for pure boron~\cite{Zhang2018b} and slightly below 
that for carbon. This is not unexpected because the K
level is deeper for elements with higher $Z$.
At temperatures above $\sim2\times10^7$ K, B$_4$C is
fully ionized and the EOS is dominated by ideal-gas 
contributions from the nuclei and the electrons.

\begin{figure}
\centering\includegraphics[width=0.42\textwidth]{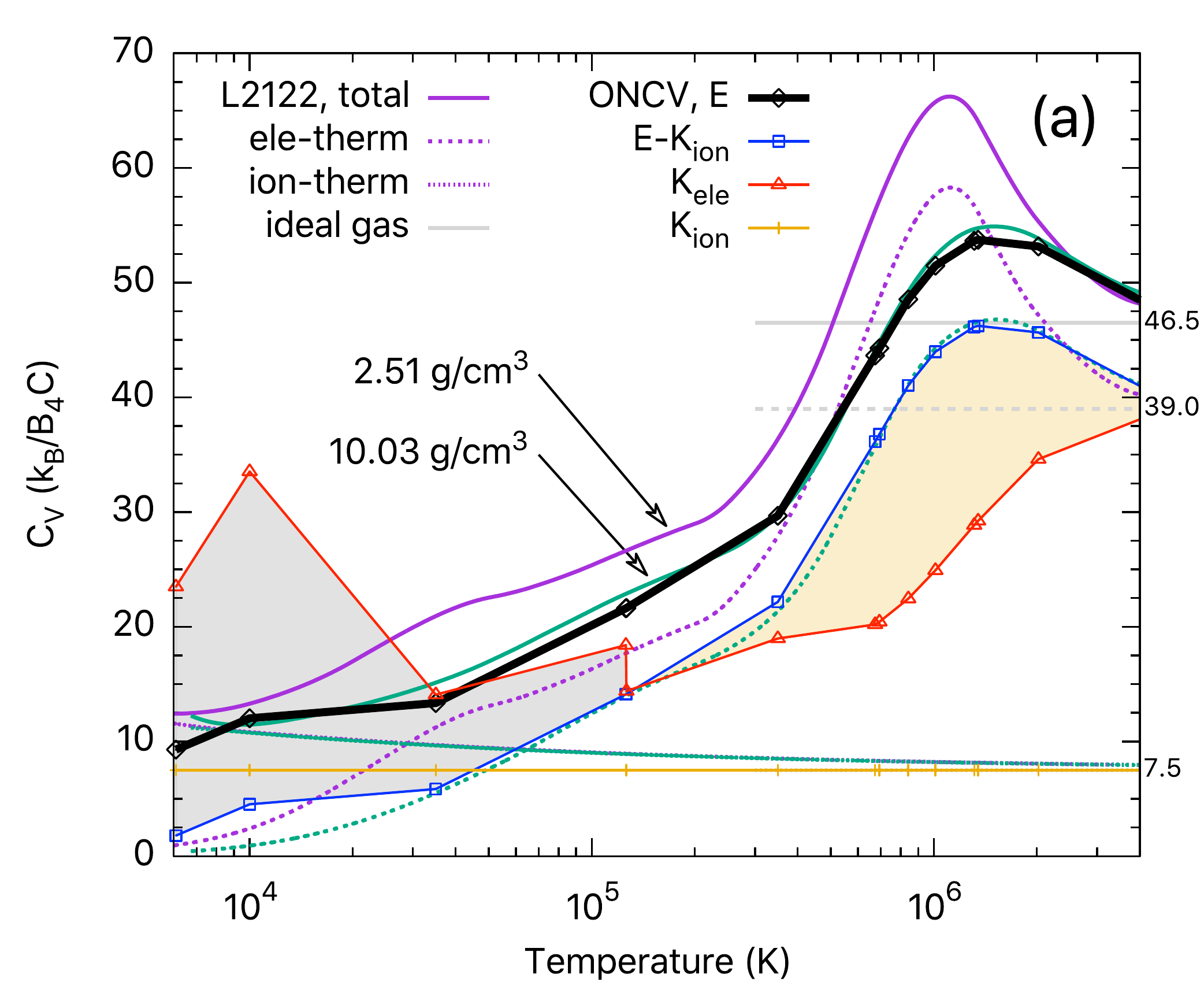}
\centering\includegraphics[width=0.42\textwidth]{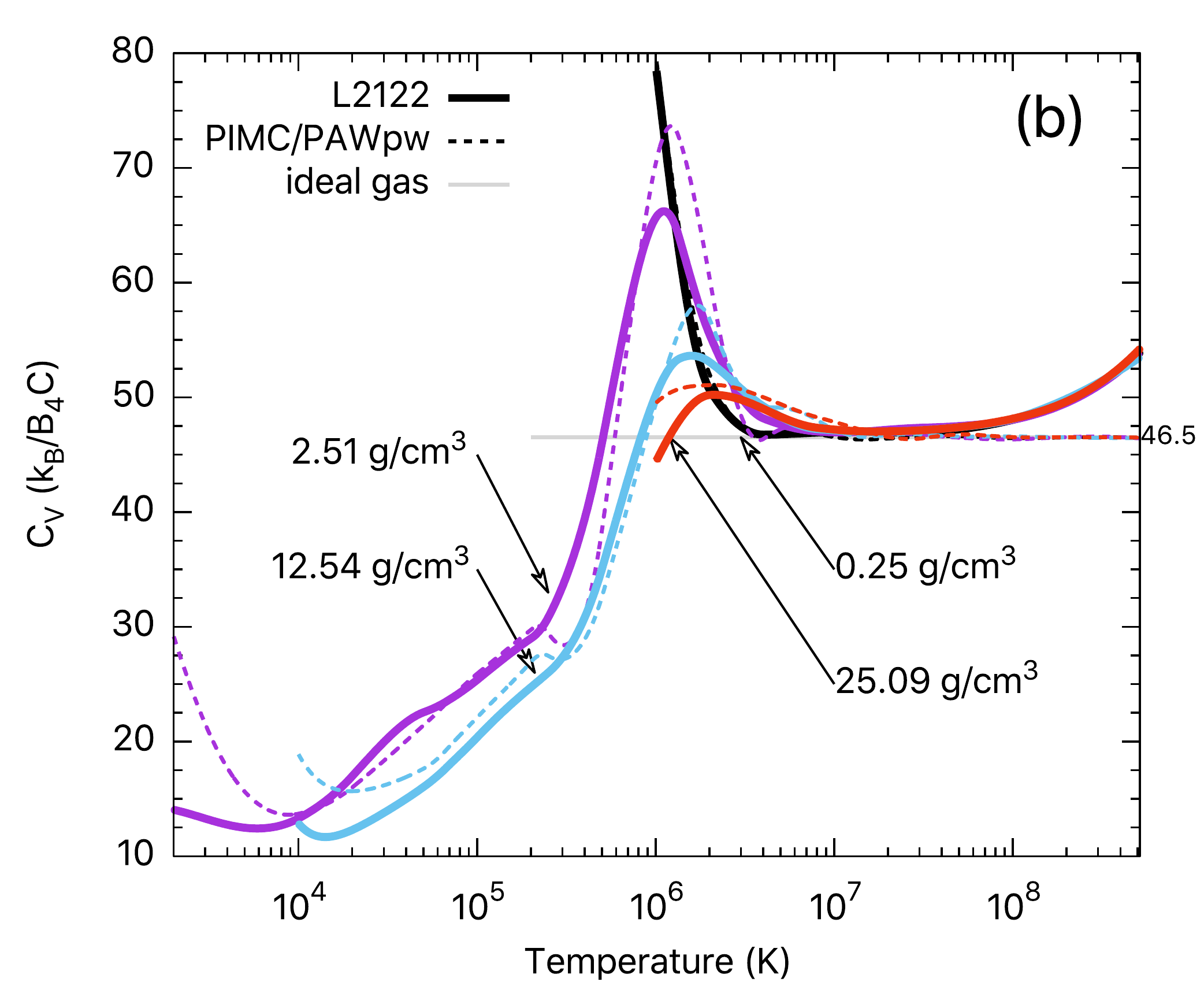}
\caption{\label{fig:cvcomp} (a) Decomposition of the heat capacity from LEOS 2122 along two isochores [2.51 g/cm$^3$ (purple) and 10.03 g/cm$^3$ (green)] and ONCV along the Hugoniot. (b) Heat capacity comparison between LEOS 2122 and PIMC/PAWpw in a broader range of temperatures. In (a), LEOS 2122 results are decomposed into electron-thermal (short dashed curves) and ion-thermal (thick dotted curves) terms. ONCV data (dark line-points) are decomposed into ion kinetic (yellow), electron kinetic (red), and interaction (i.e., all-except-ion kinetic, in blue) terms.}
\end{figure}

In order to further elucidate the roles of interaction and kinetics
in the EOS, we calculate their respective contributions to the 
heat capacity $C_\text{V}$ along the Hugoniot using the 
all-electron ONCV potential,
and the results are shown in Fig.~\ref{fig:cvcomp}(a).
The ion kinetic term ($K_\text{ion}$) contributes 7.5$k_B$/B$_4$C 
to $C_\text{V}$ independent of temperature,
where $k_\text{B}$ is the Boltzmann constant.
Electron kinetic contributions ($K_\text{ele}$) are generally 
higher (above 15$k_B$/B$_4$C)
and show two bumps, one at $10^4$ K and the other at $10^6$ K,
which can be attributed to the L- and the K-shell 
ionization, respectively. 
In contrast to $K_\text{ion}$ which follows an ideal
gas model at all temperatures, $K_\text{ele}$
is dependent on both the electronic orbitals and 
their occupancy, instead of purely on ionization,
and is not ideal gas-like until the system is fully
ionized. This can be seen from its asymptotically 
approaching the ideal gas value of 39$k_B$/B$_4$C
at above $4\times10^6$ K.

The interaction effects on the EOS are more complicated and
consist of contributions by ion-ion (``Ewald''), electron-ion (``external''),
and electron-electron (``exchange-correlation'' and ``Hartree'')
interactions. For simplicity of EOS discussions, it might
be easier to group them together than to present individually.
This is clearly shown by
the difference between the red ($K_\text{ele}$) and 
the blue ($E-K_\text{ion}$, meaning all except ion kinetic contributions)
line-points in Fig.~\ref{fig:cvcomp}(a).
The net effect of interactions can be categorized into two regions:
I (gray shaded) is below $\sim10^5$ K with $K_\text{ele}>E-K_\text{ion}$
and implying negative net contribution of interaction to $C_\text{V}$;
II (yellow shaded) is above $\sim10^5$ K with $K_\text{ele}<E-K_\text{ion}$
implying positive contributions of interaction to $C_\text{V}$.

At $\sim6\times10^3$ K, $K_\text{ele}$ contributions
are largely offset by electron-electron and ion-ion repulsion,
therefore $C_\text{V}$ is dominated by $K_\text{ion}$.
As temperature increases, the repulsive contribution
is gradually offset by the electron-ion attraction,
therefore the net interaction contribution
gradually increases to zero at $\sim1.5\times10^5$ K
and becomes positive at higher temperatures where 
K-shell ionization occurs.
At above $4\times10^6$ K, $K_\text{ele}$ and 
$K_\text{ion}$ contributions dominate
because the system is fully ionized.

Figure~\ref{fig:cvcomp}(a) also compares $C_\text{V}$ along several
isochores from L2122. As a QEOS model, L2122
decomposes the free energy into three terms:
cold-curve, ion-thermal, and electron-thermal.
The ion-thermal term (dotted lines) includes both kinetic
and interaction effects such as those from vibration.
This explains their differences relative to the K$_\text{ion}$
curves, as well as the consistency between the electron
thermal (dashed lines) and $E-K_\text{ion}$ (blue line-points),
because the cold curve does not contribute to $C_\text{V}$.
We also note [Fig.~\ref{fig:cvcomp}(b)] that the $C_\text{V}$ curves from our PIMC/PAWpw
calculations in the broad temperature range 
are consistent with L2122 predictions, except for temperatures
above $2\times10^7$ K, because the electron relativistic
effect that is included in the L2122 model raises the internal
energy and heat capacity and shifts the Hugoniot toward the
limit of 7 times compression at infinitely high temperature.

\section{Conclusions}\label{sec:conclusion}
In this work, we present a comprehensive study of the EOS of B$_4$C 
over a wide range of pressures and temperatures by implementing 
several computational methods, including PIMC, DFT-MD using standard 
plane-wave basis and PAW or ONCV pseudopotentials, ACTEX, and {\footnotesize MECCA}. 

Our EOS data by PIMC, FOE, ACTEX, and {\footnotesize MECCA} show good 
consistency at $10^6$ K where 1s electrons are ionized.
Our detailed EOS comparison provides strong evidences that
cross validate both the PIMC and the DFT-MD approaches for 
EOS studies of the partially ionized, warm-dense plasmas.

At 2.5--3.2$\times10^6$~K and 1.0--1.3$\times10^3$~Mbar,
our PIMC, {\footnotesize ACTEX}, and {\footnotesize MECCA} calculations 
uniformly predict a maximum compression of $\sim$4.55 along 
the shock Hugoniot for B$_4$C ($\rho_\text{i}$=2.51 g/cm$^3$), 
which originates from K shell ionization. 
This compression is underestimated by TF models by $\sim$0.2 (6\%).
The maximum compression ratio is similar to those of 
h-BN ($\rho_\text{i}$=2.26 g/cm$^3$)~\cite{Zhang2019bn} 
and slightly smaller than pure boron ($\rho_\text{i}$=2.31 g/cm$^3$)~\cite{Zhang2018b}.

We also report Hugoniot data up to $\sim 61$ Mbar from experiments 
at the NIF. 
The measured data show good agreement with our theoretical predictions based on DFT-MD.

By comparing QEOS models with the electron thermal term constructed in 
different ways (Purgatorio in LEOS 2122 or TF in LEOS 2120/SESAME 7082),
we find that the Purgatorio-based EOS models provide excellent overall agreement 
with our numerical simulations, similar to our previous studies on pure boron and BN.
Because the largest differences in the Hugoniot response of the models occurs 
near peak compression, performing experiments for materials near peak compression~\cite{Swift2012,Kritcher2016,Nilsen2016,Swift2018,Doppner2018} would provide a rigorous experimental test of our understanding 
of electronic structure in high energy density plasmas. 
It would also be worthwhile to pursue experiments that provide measurements of 
the temperature and the pressure in either 
Hugoniot or off-Hugoniot experiments, which would provide data to 
test the first principle calculations.

Based on the experimental data, we have developed three new EOS models (L2123, L2124, and L2125) by variations of the ion thermal EOS model to span the range of experimental error bars. 
These models were developed to span the range of
EOS models that are consistent with the experimental error bars.  
1D hydrodynamic simulations of direct drive implosions with a B$_4$C ablator demonstrate that the nominal
polar direct drive exploding pusher design is not sensitive to the equation of state model.  Our work should motivate similar studies for future ICF designs using B$_4$C ablators. 

\section{Supplementary Material}
See the supplementary material (Table~\ref{tab:eostable}) for the EOS data table of B$_4$C from this study.

\begin{acknowledgments}  
This work was in part performed under the auspices of the U.S. Department of Energy by Lawrence Livermore National Laboratory under Contract No. DE-AC52-07NA27344.
Computational support was provided by LLNL high-performance
computing facility (Quartz) and the Blue Waters 
sustained-petascale computing project (NSF ACI 1640776).
B.M. is supported by the U. S. Department of Energy (grant DE-SC0016248) and the University of California. S.Z. is partially supported by the PLS-Postdoctoral Grant of LLNL. 

This document was prepared as an account of work sponsored by an agency of the United States government. Neither the United States government nor any agency thereof, nor any of their employees,
makes any warranty, express or implied, or assumes any legal
liability or responsibility for the accuracy, completeness, or
usefulness of any information, apparatus, product, or process
disclosed, or represents that its use would not infringe privately owned rights. Reference herein to any specific commercial product, process, or service by trade name, trademark,
manufacturer, or otherwise does not necessarily constitute
or imply its endorsement, recommendation, or favoring by
the U.S. Government or any agency thereof. The views and
opinions of authors expressed herein do not necessarily state
or reflect those of the U.S. Government or any agency thereof, and shall not be used for advertising or product endorsement purposes.
\end{acknowledgments}


\begin{longtable*}{ccccccl}
\caption{\label{tab:eostable} {\bf Supplementary Material}: first-principles equation of state data for B$_4$C based on PIMC and DFT-MD [PAWpw, ONCVpw(1.07), FOE/ae-pw(2.06)] simulations by Burkhard Militzer, Shuai Zhang and Lin H. Yang. Energies are relative to the corresponding values of B$_4$C at ambient condition (300 K, 2.51 g/cc).}\\
\hline
$\rho$ & $T$ & $P$ & $P_\text{error}$ & $E$ & $E_\text{error}$  & note \\
(g/cm$^3$) & (K) & (GPa) & (GPa) & (Ha/B$_4$C) & (Ha/B$_4$C)  &  \\
\hline 
\endfirsthead
\multicolumn{7}{@{}l}{\ldots continued}\\
\hline
$\rho$ & $T$ & $P$ & $P_\text{error}$ & $E$ & $E_\text{error}$  & note \\
(g/cm$^3$) & (K) & (GPa) & (GPa) & (Ha/B$_4$C) & (Ha/B$_4$C)  &  \\
\hline
\endhead 
\hline
\multicolumn{7}{r@{}}{continued \ldots}\\
\endfoot
\hline
\endlastfoot
  0.2509 &    1010479 &        978.3 &      9.7 &     206.906 &  1.219  &   PIMC\_A3368 \\
  0.2509 &    1347305 &       1402.8 &      9.4 &     283.310 &  1.184  &   PIMC\_A3367 \\
  0.2509 &    2020958 &       2254.4 &     10.9 &     404.807 &  1.372  &   PIMC\_A3366 \\
  0.2509 &    4041916 &       4641.0 &     15.3 &     711.485 &  1.926  &   PIMC\_A3365 \\
  0.2509 &    8083831 &       9400.4 &     18.0 &    1312.607 &  2.266  &   PIMC\_A3364 \\
  0.2509 &   16167663 &      18839.9 &     20.1 &    2501.695 &  2.528  &   PIMC\_A3363 \\
  0.2509 &   32335325 &      37744.2 &     28.3 &    4881.135 &  3.559  &   PIMC\_A3362 \\
  0.2509 &   64670651 &      75667.1 &     38.4 &    9653.671 &  4.832  &   PIMC\_A3361 \\
  0.2509 &  129341301 &     151298.9 &     50.4 &   19170.798 &  6.348  &   PIMC\_A3360 \\
  0.2509 &  258682602 &     302675.0 &     70.6 &   38219.473 &  8.871  &   PIMC\_A3359 \\
  0.2509 &  517365204 &     605430.7 &     65.9 &   76316.150 &  8.280  &   PIMC\_A3358 \\
  0.6272 &    1347305 &       3476.4 &     68.1 &     266.271 &  3.426  &   PIMC\_A3565 \\
  0.6272 &    2020958 &       5653.7 &     69.0 &     395.583 &  3.473  &   PIMC\_A3564 \\
  0.6272 &    4041916 &      11487.2 &     59.9 &     700.416 &  3.015  &   PIMC\_A3563 \\
  0.6272 &    8083831 &      23286.6 &     79.8 &    1298.108 &  4.016  &   PIMC\_A3562 \\
  0.6272 &   16167663 &      47143.0 &     85.3 &    2500.912 &  4.296  &   PIMC\_A3561 \\
  0.6272 &   32335325 &      94249.1 &    112.8 &    4873.197 &  5.674  &   PIMC\_A3560 \\
  0.6272 &   64670651 &     188925.7 &    154.8 &    9639.451 &  7.786  &   PIMC\_A3559 \\
  0.6272 &  129341301 &     378160.5 &    174.2 &   19164.543 &  8.762  &   PIMC\_A3558 \\
  0.6272 &  258682602 &     756073.9 &    270.9 &   38186.621 & 13.628  &   PIMC\_A3557 \\
  0.6272 &  517365204 &    1513078.5 &    253.9 &   76289.408 & 12.782  &   PIMC\_A3556 \\
  1.2544 &    1347305 &       6449.7 &    134.8 &     238.128 &  3.390  &   PIMC\_A3576 \\
  1.2544 &    2020958 &      11170.0 &    140.0 &     381.263 &  3.522  &   PIMC\_A3575 \\
  1.2544 &    4041916 &      22672.4 &    119.1 &     686.744 &  3.000  &   PIMC\_A3574 \\
  1.2544 &    8083831 &      46886.6 &    158.6 &    1301.944 &  3.988  &   PIMC\_A3573 \\
  1.2544 &   16167663 &      94533.2 &    157.3 &    2504.194 &  3.960  &   PIMC\_A3572 \\
  1.2544 &   32335325 &     188899.6 &    214.1 &    4880.955 &  5.392  &   PIMC\_A3571 \\
  1.2544 &   64670651 &     378262.2 &    264.7 &    9647.721 &  6.666  &   PIMC\_A3570 \\
  1.2544 &  129341301 &     757312.8 &    382.3 &   19187.998 &  9.627  &   PIMC\_A3569 \\
  1.2544 &  258682602 &    1514056.9 &    533.5 &   38232.909 & 13.454  &   PIMC\_A3568 \\
  1.2544 &  517365204 &    3026294.1 &    690.7 &   76291.346 & 17.394  &   PIMC\_A3567 \\
  2.5087 &       2000 &          8.8 &      0.3 &       0.178 &  0.001  &   PBE\_B4C120\_0137 \\
  2.5087 &       6736 &         34.2 &      0.2 &       0.451 &  0.000  &   PBE\_B4C120\_0138 \\
  2.5087 &      10000 &         49.7 &      0.2 &       0.593 &  0.001  &   PBE\_B4C120\_0139 \\
  2.5087 &      20000 &         98.2 &      0.2 &       1.060 &  0.001  &   PBE\_B4C120\_0140 \\
  2.5087 &      50523 &        264.2 &      0.4 &       2.906 &  0.001  &   PBE\_B4C120\_0141 \\
  2.5087 &      67364 &        360.7 &      0.5 &       4.102 &  0.002  &   PBE\_B4C120\_0142 \\
  2.5087 &     101047 &        568.4 &      0.7 &       6.747 &  0.003  &   PBE\_B4C120\_0143 \\
  2.5087 &     202095 &       1241.3 &      1.0 &      15.738 &  0.010  &   PBE\_B4C120\_0145 \\
  2.5087 &     252619 &       1590.9 &      1.5 &      20.516 &  0.012  &   PBE\_B4C120\_0146 \\
  2.5087 &     505239 &       3447.0 &      1.4 &      45.647 &  0.020  &   PBE\_B4C30\_0019 \\
  2.5087 &    1347305 &      12396.6 &    265.9 &     215.509 &  3.347  &   PIMC\_A3587 \\
  2.5087 &    2020958 &      21582.0 &    288.2 &     357.506 &  3.625  &   PIMC\_A3586 \\
  2.5087 &    4041916 &      45205.4 &    248.0 &     676.904 &  3.119  &   PIMC\_A3585 \\
  2.5087 &    8083831 &      92769.4 &    310.1 &    1283.561 &  3.902  &   PIMC\_A3584 \\
  2.5087 &   16167663 &     187865.4 &    337.2 &    2484.609 &  4.243  &   PIMC\_A3583 \\
  2.5087 &   32335325 &     378004.1 &    412.5 &    4880.292 &  5.206  &   PIMC\_A3582 \\
  2.5087 &   64670651 &     756470.6 &    612.0 &    9644.225 &  7.704  &   PIMC\_A3581 \\
  2.5087 &  129341301 &    1510913.9 &    833.1 &   19138.679 & 10.495  &   PIMC\_A3580 \\
  2.5087 &  258682602 &    3024944.4 &   1072.2 &   38190.873 & 13.480  &   PIMC\_A3579 \\
  2.5087 &  517365204 &    6054012.4 &   1131.2 &   76307.047 & 14.213  &   PIMC\_A3578 \\
  5.0174 &    1347305 &      25101.2 &    503.4 &     200.936 &  3.166  &   PIMC\_A3598 \\
  5.0174 &    2020958 &      38909.9 &    550.8 &     313.556 &  3.472  &   PIMC\_A3597 \\
  5.0174 &    4041916 &      88884.2 &    486.5 &     655.812 &  3.062  &   PIMC\_A3596 \\
  5.0174 &    8083831 &     185650.7 &    597.0 &    1276.586 &  3.755  &   PIMC\_A3595 \\
  5.0174 &   16167663 &     374783.7 &    638.2 &    2472.912 &  4.010  &   PIMC\_A3594 \\
  5.0174 &   32335325 &     753018.7 &    890.7 &    4856.619 &  5.603  &   PIMC\_A3593 \\
  5.0174 &   64670651 &    1511237.5 &   1126.5 &    9629.535 &  7.068  &   PIMC\_A3592 \\
  5.0174 &  129341301 &    3027710.6 &   1432.9 &   19172.825 &  8.963  &   PIMC\_A3591 \\
  5.0174 &  258682602 &    6056069.4 &   2235.2 &   38226.650 & 14.063  &   PIMC\_A3590 \\
  5.0174 &  517365204 &   12103991.4 &   2778.3 &   76278.717 & 17.526  &   PIMC\_A3589 \\
  5.0175 &       6736 &        443.7 &      0.3 &       0.892 &  0.001  &   PBE\_B4C120\_0148 \\
  5.0175 &      10000 &        478.5 &      0.4 &       1.042 &  0.001  &   PBE\_B4C120\_0149 \\
  5.0175 &      20000 &        580.7 &      0.8 &       1.504 &  0.002  &   PBE\_B4C120\_0150 \\
  5.0175 &      50523 &        908.6 &      1.0 &       3.216 &  0.003  &   PBE\_B4C120\_0151 \\
  5.0175 &      67364 &       1106.9 &      1.1 &       4.344 &  0.004  &   PBE\_B4C120\_0152 \\
  5.0175 &     101047 &       1516.6 &      1.5 &       6.821 &  0.005  &   PBE\_B4C120\_0153 \\
  5.0175 &     126309 &       1838.0 &      0.9 &       8.832 &  0.003  &   PBE\_B4C120\_0154 \\
  5.0175 &     202095 &       2844.7 &      0.9 &      15.336 &  0.004  &   PBE\_B4C120\_0155 \\
  5.0175 &     252619 &       3535.9 &      2.1 &      19.938 &  0.008  &   PBE\_B4C120\_0156 \\
  5.0175 &     505239 &       7122.1 &      3.9 &      44.159 &  0.029  &   PBE\_B4C30\_0035 \\
  7.5261 &    1010479 &      24092.6 &    676.4 &     119.081 &  2.836  &   PIMC\_A3610 \\
  7.5261 &    1347305 &      34984.9 &    616.5 &     180.443 &  2.586  &   PIMC\_A3609 \\
  7.5261 &    2020958 &      58336.2 &    713.7 &     303.149 &  2.991  &   PIMC\_A3608 \\
  7.5261 &    4041916 &     132474.5 &    725.5 &     643.917 &  3.038  &   PIMC\_A3607 \\
  7.5261 &    8083831 &     278394.0 &    933.9 &    1270.287 &  3.920  &   PIMC\_A3606 \\
  7.5261 &   16167663 &     561169.8 &    988.3 &    2464.149 &  4.144  &   PIMC\_A3605 \\
  7.5261 &   32335325 &    1132212.6 &   1304.3 &    4864.353 &  5.483  &   PIMC\_A3604 \\
  7.5261 &   64670651 &    2266120.7 &   1450.6 &    9624.077 &  6.105  &   PIMC\_A3603 \\
  7.5261 &  129341301 &    4537038.6 &   2389.5 &   19151.236 &  9.991  &   PIMC\_A3602 \\
  7.5261 &  258682602 &    9075883.8 &   3734.0 &   38189.980 & 15.688  &   PIMC\_A3601 \\
  7.5261 &  517365204 &   18160616.9 &   3508.9 &   76296.115 & 14.737  &   PIMC\_A3600 \\
  7.5262 &       6736 &       1336.5 &      0.5 &       1.866 &  0.001  &   PBE\_B4C120\_0158 \\
  7.5262 &      10000 &       1391.2 &      0.5 &       2.027 &  0.001  &   PBE\_B4C120\_0159 \\
  7.5262 &      20000 &       1545.6 &      0.8 &       2.501 &  0.001  &   PBE\_B4C120\_0160 \\
  7.5262 &      50523 &       2017.5 &      0.8 &       4.148 &  0.002  &   PBE\_B4C120\_0161 \\
  7.5262 &      67364 &       2297.1 &      1.4 &       5.208 &  0.003  &   PBE\_B4C120\_0162 \\
  7.5262 &     101047 &       2894.1 &      2.0 &       7.583 &  0.006  &   PBE\_B4C120\_0163 \\
  7.5262 &     126309 &       3372.8 &      2.5 &       9.535 &  0.007  &   PBE\_B4C120\_0164 \\
  7.5262 &     202095 &       4850.5 &      1.8 &      15.857 &  0.006  &   PBE\_B4C120\_0165 \\
  7.5262 &     252619 &       5870.5 &      1.9 &      20.376 &  0.006  &   PBE\_B4C120\_0166 \\
  7.5262 &     505239 &      11135.2 &      7.6 &      44.177 &  0.036  &   PBE\_B4C30\_0051 \\
 10.0349 &       6736 &       2699.3 &      0.6 &       3.105 &  0.001  &   PBE\_B4C120\_0168 \\
 10.0349 &      10000 &       2778.3 &      0.7 &       3.276 &  0.001  &   PBE\_B4C120\_0169 \\
 10.0349 &      20000 &       2989.2 &      1.1 &       3.770 &  0.002  &   PBE\_B4C120\_0170 \\
 10.0349 &      50523 &       3597.4 &      2.7 &       5.382 &  0.005  &   PBE\_B4C120\_0171 \\
 10.0349 &      67364 &       3954.4 &      3.0 &       6.403 &  0.006  &   PBE\_B4C120\_0172 \\
 10.0349 &     101047 &       4723.7 &      3.1 &       8.694 &  0.007  &   PBE\_B4C120\_0173 \\
 10.0349 &     126309 &       5332.8 &      2.9 &      10.572 &  0.006  &   PBE\_B4C120\_0174 \\
 10.0349 &     202095 &       7267.9 &      3.5 &      16.787 &  0.009  &   PBE\_B4C120\_0175 \\
 10.0349 &     252619 &       8592.0 &      4.0 &      21.239 &  0.010  &   PBE\_B4C120\_0176 \\
 10.0349 &     505239 &      15437.1 &      9.6 &      44.666 &  0.036  &   PBE\_B4C30\_0067 \\
 10.0349 &    1010479 &      34986.5 &   1044.5 &     123.038 &  3.285  &   PIMC\_A3621 \\
 10.0349 &    1347305 &      46667.0 &   1012.2 &     173.850 &  3.185  &   PIMC\_A3620 \\
 10.0349 &    2020958 &      77656.4 &   1082.6 &     294.180 &  3.405  &   PIMC\_A3619 \\
 10.0349 &    4041916 &     177948.6 &    977.7 &     641.285 &  3.079  &   PIMC\_A3618 \\
 10.0349 &    8083831 &     367611.3 &   1269.2 &    1254.130 &  3.992  &   PIMC\_A3617 \\
 10.0349 &   16167663 &     746834.2 &   1262.5 &    2456.294 &  3.969  &   PIMC\_A3616 \\
 10.0349 &   32335325 &    1508100.1 &   1689.9 &    4857.032 &  5.321  &   PIMC\_A3615 \\
 10.0349 &   64670651 &    3023617.9 &   2169.2 &    9628.178 &  6.870  &   PIMC\_A3614 \\
 10.0349 &  129341301 &    6043779.5 &   2638.7 &   19132.188 &  8.317  &   PIMC\_A3613 \\
 10.0349 &  258682602 &   12096112.7 &   4951.6 &   38172.125 & 15.585  &   PIMC\_A3612 \\
 10.0349 &  517365204 &   24216797.6 &   5382.9 &   76302.638 & 16.961  &   PIMC\_A3611 \\
 11.2892 &    1010479 &      41709.7 &   1143.1 &     127.812 &  3.197  &   PIMC\_A3632 \\
 11.2892 &    1347305 &      54706.0 &   1098.7 &     177.545 &  3.072  &   PIMC\_A3631 \\
 11.2892 &    2020958 &      84970.2 &   1195.4 &     284.316 &  3.347  &   PIMC\_A3630 \\
 11.2892 &    4041916 &     194649.0 &   1080.3 &     622.685 &  3.019  &   PIMC\_A3629 \\
 11.2892 &    8083831 &     414955.6 &   1405.5 &    1255.831 &  3.931  &   PIMC\_A3628 \\
 11.2892 &   16167663 &     838384.2 &   1474.3 &    2449.395 &  4.124  &   PIMC\_A3627 \\
 11.2892 &   32335325 &    1692733.1 &   2027.9 &    4844.705 &  5.676  &   PIMC\_A3626 \\
 11.2892 &   64670651 &    3407184.7 &   2353.1 &    9643.450 &  6.580  &   PIMC\_A3625 \\
 11.2892 &  129341301 &    6802116.0 &   4041.0 &   19138.514 & 11.329  &   PIMC\_A3624 \\
 11.2892 &  258682602 &   13620367.2 &   4405.1 &   38206.034 & 12.316  &   PIMC\_A3623 \\
 11.2892 &  517365204 &   27241435.2 &   5301.2 &   76295.358 & 14.821  &   PIMC\_A3622 \\
 12.5436 &      10000 &       4628.5 &      0.7 &       4.676 &  0.001  &   PBE\_B4C120\_0179 \\
 12.5436 &      20000 &       4901.3 &      1.2 &       5.196 &  0.002  &   PBE\_B4C120\_0180 \\
 12.5436 &      50523 &       5647.4 &      1.8 &       6.797 &  0.003  &   PBE\_B4C120\_0181 \\
 12.5436 &      67364 &       6070.7 &      1.0 &       7.780 &  0.002  &   PBE\_B4C120\_0182 \\
 12.5436 &     101047 &       6998.4 &      2.2 &      10.002 &  0.005  &   PBE\_B4C120\_0183 \\
 12.5436 &     126309 &       7737.4 &      5.0 &      11.839 &  0.010  &   PBE\_B4C120\_0184 \\
 12.5436 &     202095 &      10085.9 &      4.5 &      17.947 &  0.011  &   PBE\_B4C120\_0185 \\
 12.5436 &     252619 &      11697.0 &      4.5 &      22.338 &  0.011  &   PBE\_B4C120\_0186 \\
 12.5436 &     505239 &      20042.1 &      7.9 &      45.469 &  0.029  &   PBE\_B4C30\_0083 \\
 12.5436 &    1347305 &      58420.8 &   1183.6 &     169.416 &  2.978  &   PIMC\_A3642 \\
 12.5436 &    2020958 &      98531.3 &   1292.6 &     291.848 &  3.249  &   PIMC\_A3641 \\
 12.5436 &    4041916 &     216565.9 &   1198.0 &     620.836 &  3.019  &   PIMC\_A3640 \\
 12.5436 &    8083831 &     457590.9 &   1556.2 &    1244.895 &  3.923  &   PIMC\_A3639 \\
 12.5436 &   16167663 &     930693.5 &   1648.7 &    2445.848 &  4.151  &   PIMC\_A3638 \\
 12.5436 &   32335325 &    1879207.6 &   2326.5 &    4839.516 &  5.857  &   PIMC\_A3637 \\
 12.5436 &   64670651 &    3774359.1 &   3080.0 &    9613.114 &  7.730  &   PIMC\_A3636 \\
 12.5436 &  129341301 &    7564430.1 &   3763.7 &   19153.993 &  9.483  &   PIMC\_A3635 \\
 12.5436 &  258682602 &   15135191.9 &   5628.8 &   38208.688 & 14.165  &   PIMC\_A3634 \\
 12.5436 &  517365204 &   30257769.9 &   4928.2 &   76268.123 & 12.396  &   PIMC\_A3633 \\
 15.0523 &    1010479 &      53347.1 &   1465.9 &     117.897 &  3.077  &   PIMC\_A3654 \\
 15.0523 &    1347305 &      73340.1 &   1428.0 &     171.417 &  2.997  &   PIMC\_A3653 \\
 15.0523 &    2020958 &     119085.1 &   1610.1 &     288.524 &  3.371  &   PIMC\_A3652 \\
 15.0523 &    4041916 &     262569.8 &   1486.4 &     621.390 &  3.115  &   PIMC\_A3651 \\
 15.0523 &    8083831 &     552139.3 &   1882.7 &    1247.669 &  3.949  &   PIMC\_A3650 \\
 15.0523 &   16167663 &    1117688.5 &   1965.7 &    2444.803 &  4.127  &   PIMC\_A3649 \\
 15.0523 &   32335325 &    2260439.8 &   2810.9 &    4848.438 &  5.904  &   PIMC\_A3648 \\
 15.0523 &   64670651 &    4537668.1 &   3241.8 &    9629.701 &  6.808  &   PIMC\_A3647 \\
 15.0523 &  129341301 &    9069639.2 &   5110.8 &   19136.901 & 10.681  &   PIMC\_A3646 \\
 15.0523 &  258682602 &   18144450.8 &   7854.1 &   38170.075 & 16.462  &   PIMC\_A3645 \\
 15.0523 &  517365204 &   36320053.4 &   6770.2 &   76289.249 & 14.184  &   PIMC\_A3644 \\
 15.0524 &       6736 &       6775.1 &     12.0 &       5.933 &  0.017  &   PBE\_B4C120\_0188 \\
 15.0524 &      20000 &       7261.8 &      1.1 &       6.704 &  0.001  &   PBE\_B4C120\_0190 \\
 15.0524 &      50523 &       8149.9 &      1.9 &       8.315 &  0.002  &   PBE\_B4C120\_0191 \\
 15.0524 &      67364 &       8648.5 &      2.2 &       9.287 &  0.004  &   PBE\_B4C120\_0192 \\
 15.0524 &     101047 &       9713.2 &      3.5 &      11.446 &  0.005  &   PBE\_B4C120\_0193 \\
 15.0524 &     126309 &      10560.4 &      3.8 &      13.222 &  0.007  &   PBE\_B4C120\_0194 \\
 15.0524 &     202095 &      13290.7 &      5.6 &      19.243 &  0.011  &   PBE\_B4C120\_0195 \\
 15.0524 &     252619 &      15162.4 &      6.5 &      23.602 &  0.016  &   PBE\_B4C120\_0196 \\
 15.0524 &     505239 &      24961.7 &      9.4 &      46.532 &  0.025  &   PBE\_B4C30\_0099 \\
 17.5610 &    1010479 &      62336.2 &   1580.2 &     115.857 &  2.836  &   PIMC\_A3665 \\
 17.5610 &    1347305 &      85805.3 &   1608.8 &     168.247 &  2.891  &   PIMC\_A3664 \\
 17.5610 &    2020958 &     136741.9 &   1818.2 &     279.581 &  3.273  &   PIMC\_A3663 \\
 17.5610 &    4041916 &     300880.5 &   1708.3 &     607.206 &  3.071  &   PIMC\_A3662 \\
 17.5610 &    8083831 &     641212.4 &   2159.3 &    1238.969 &  3.886  &   PIMC\_A3661 \\
 17.5610 &   16167663 &    1305330.3 &   2254.3 &    2444.524 &  4.053  &   PIMC\_A3660 \\
 17.5610 &   32335325 &    2628791.0 &   3092.4 &    4831.496 &  5.553  &   PIMC\_A3659 \\
 17.5610 &   64670651 &    5290724.2 &   3841.7 &    9621.624 &  6.923  &   PIMC\_A3658 \\
 17.5610 &  129341301 &   10568812.8 &   5871.5 &   19112.975 & 10.583  &   PIMC\_A3657 \\
 17.5610 &  258682602 &   21175001.0 &   8164.4 &   38180.974 & 14.699  &   PIMC\_A3656 \\
 17.5610 &  517365204 &   42363812.8 &   8271.6 &   76271.245 & 14.867  &   PIMC\_A3655 \\
 17.5611 &      10000 &       9598.7 &     13.4 &       7.613 &  0.019  &   PBE\_B4C120\_0199 \\
 17.5611 &      20000 &      10070.4 &      1.8 &       8.279 &  0.002  &   PBE\_B4C120\_0200 \\
 17.5611 &      50523 &      11094.5 &      1.8 &       9.900 &  0.002  &   PBE\_B4C120\_0201 \\
 17.5611 &      67364 &      11661.7 &      2.9 &      10.860 &  0.004  &   PBE\_B4C120\_0202 \\
 17.5611 &     101047 &      12849.0 &      3.1 &      12.955 &  0.005  &   PBE\_B4C120\_0203 \\
 17.5611 &     126309 &      13818.2 &      4.7 &      14.720 &  0.007  &   PBE\_B4C120\_0204 \\
 17.5611 &     202095 &      16870.0 &      6.6 &      20.642 &  0.011  &   PBE\_B4C120\_0205 \\
 17.5611 &     252619 &      18989.2 &      6.6 &      24.999 &  0.016  &   PBE\_B4C120\_0206 \\
 17.5611 &     505239 &      30145.1 &     15.9 &      47.653 &  0.046  &   PBE\_B4C30\_0115 \\
 20.0697 &    1010479 &      70289.2 &   1815.5 &     111.781 &  2.854  &   PIMC\_A3676 \\
 20.0697 &    1347305 &      93427.7 &   1819.3 &     158.288 &  2.859  &   PIMC\_A3675 \\
 20.0697 &    2020958 &     159835.5 &   2151.1 &     281.486 &  3.384  &   PIMC\_A3674 \\
 20.0697 &    4041916 &     341929.4 &   2005.1 &     599.969 &  3.152  &   PIMC\_A3673 \\
 20.0697 &    8083831 &     725349.4 &   2370.9 &    1223.976 &  3.732  &   PIMC\_A3672 \\
 20.0697 &   16167663 &    1489538.2 &   2722.7 &    2438.785 &  4.282  &   PIMC\_A3671 \\
 20.0697 &   32335325 &    3008323.1 &   3484.4 &    4835.791 &  5.471  &   PIMC\_A3670 \\
 20.0697 &   64670651 &    6033813.2 &   4696.1 &    9600.594 &  7.390  &   PIMC\_A3669 \\
 20.0697 &  129341301 &   12085003.6 &   6140.8 &   19121.899 &  9.673  &   PIMC\_A3668 \\
 20.0697 &  258682602 &   24204304.1 &   9811.3 &   38186.636 & 15.483  &   PIMC\_A3667 \\
 20.0697 &  517365204 &   48408536.7 &   9345.6 &   76258.850 & 14.675  &   PIMC\_A3666 \\
 20.0698 &      10000 &      12762.8 &      0.9 &       9.181 &  0.001  &   PBE\_B4C120\_0209 \\
 20.0698 &      20000 &      13320.4 &      1.4 &       9.905 &  0.001  &   PBE\_B4C120\_0210 \\
 20.0698 &      50523 &      14477.6 &      2.4 &      11.543 &  0.003  &   PBE\_B4C120\_0211 \\
 20.0698 &      67364 &      15100.8 &      4.8 &      12.485 &  0.006  &   PBE\_B4C120\_0212 \\
 20.0698 &     101047 &      16425.3 &      3.8 &      14.561 &  0.006  &   PBE\_B4C120\_0213 \\
 20.0698 &     126309 &      17466.7 &      4.6 &      16.264 &  0.008  &   PBE\_B4C120\_0214 \\
 20.0698 &     202095 &      20828.9 &      5.1 &      22.165 &  0.009  &   PBE\_B4C120\_0215 \\
 20.0698 &     252619 &      23147.8 &      6.6 &      26.468 &  0.013  &   PBE\_B4C120\_0216 \\
 20.0698 &     505239 &      35571.3 &     14.3 &      48.863 &  0.030  &   PBE\_B4C30\_0131 \\
 25.0872 &    1010479 &      87214.1 &   2216.6 &     107.373 &  2.787  &   PIMC\_A3687 \\
 25.0872 &    1347305 &     123235.7 &   2239.0 &     160.989 &  2.820  &   PIMC\_A3686 \\
 25.0872 &    2020958 &     196124.9 &   2664.1 &     269.727 &  3.354  &   PIMC\_A3685 \\
 25.0872 &    4041916 &     428037.6 &   2444.2 &     593.804 &  3.077  &   PIMC\_A3684 \\
 25.0872 &    8083831 &     908550.6 &   2946.7 &    1220.664 &  3.706  &   PIMC\_A3683 \\
 25.0872 &   16167663 &    1864015.2 &   3317.5 &    2436.957 &  4.176  &   PIMC\_A3682 \\
 25.0872 &   32335325 &    3747985.1 &   4716.4 &    4816.846 &  5.943  &   PIMC\_A3681 \\
 25.0872 &   64670651 &    7543547.7 &   5674.7 &    9599.259 &  7.151  &   PIMC\_A3680 \\
 25.0872 &  129341301 &   15119588.9 &   6727.7 &   19136.352 &  8.486  &   PIMC\_A3679 \\
 25.0872 &  258682602 &   30258390.7 &  11453.7 &   38187.745 & 14.433  &   PIMC\_A3678 \\
 25.0872 &  517365204 &   60534837.3 &  10922.3 &   76287.187 & 13.773  &   PIMC\_A3677 \\
 30.1046 &    1010479 &     107601.8 &   2604.2 &     107.318 &  2.730  &   PIMC\_A3698 \\
 30.1046 &    1347305 &     140893.9 &   2693.8 &     149.817 &  2.828  &   PIMC\_A3697 \\
 30.1046 &    2020958 &     232091.0 &   3088.5 &     261.337 &  3.240  &   PIMC\_A3696 \\
 30.1046 &    4041916 &     512012.4 &   2940.2 &     586.090 &  3.084  &   PIMC\_A3695 \\
 30.1046 &    8083831 &    1078105.1 &   3821.0 &    1202.934 &  4.006  &   PIMC\_A3694 \\
 30.1046 &   16167663 &    2228993.3 &   3817.0 &    2424.901 &  4.009  &   PIMC\_A3693 \\
 30.1046 &   32335325 &    4520030.4 &   5136.8 &    4837.597 &  5.372  &   PIMC\_A3692 \\
 30.1046 &   64670651 &    9058273.3 &   7317.4 &    9603.443 &  7.667  &   PIMC\_A3691 \\
 30.1046 &  129341301 &   18123765.5 &   9637.7 &   19113.151 & 10.120  &   PIMC\_A3690 \\
 30.1046 &  258682602 &   36320568.3 &  12420.2 &   38198.164 & 13.042  &   PIMC\_A3689 \\
 30.1046 &  517365204 &   72622107.5 &  13340.5 &   76265.622 & 14.002  &   PIMC\_A3688 \\
 37.6307 &    1010479 &     147119.9 &   3181.3 &     113.815 &  2.671  &   PIMC\_A3709 \\
 37.6307 &    1347305 &     193405.1 &   3225.9 &     159.107 &  2.707  &   PIMC\_A3708 \\
 37.6307 &    2020958 &     298300.7 &   3899.0 &     261.283 &  3.272  &   PIMC\_A3707 \\
 37.6307 &    4041916 &     636287.4 &   3753.1 &     574.957 &  3.146  &   PIMC\_A3706 \\
 37.6307 &    8083831 &    1348937.5 &   4607.3 &    1197.395 &  3.863  &   PIMC\_A3705 \\
 37.6307 &   16167663 &    2777583.7 &   4929.8 &    2412.462 &  4.135  &   PIMC\_A3704 \\
 37.6307 &   32335325 &    5640862.8 &   6172.6 &    4825.974 &  5.175  &   PIMC\_A3703 \\
 37.6307 &   64670651 &   11304155.2 &   8029.8 &    9584.400 &  6.776  &   PIMC\_A3702 \\
 37.6307 &  129341301 &   22665528.6 &  10246.9 &   19119.821 &  8.547  &   PIMC\_A3701 \\
 37.6307 &  258682602 &   45377914.9 &  14656.3 &   38176.243 & 12.356  &   PIMC\_A3700 \\
 37.6307 &  517365204 &   90789135.2 &  18077.2 &   76272.864 & 15.166  &   PIMC\_A3699 \\
 50.1743 &    1010479 &     228810.2 &   3608.3 &     129.142 &  2.269  &   PIMC\_A3379 \\
 50.1743 &    1347305 &     268222.6 &   3711.4 &     158.940 &  2.338  &   PIMC\_A3378 \\
 50.1743 &    2020958 &     416612.9 &   4532.8 &     264.081 &  2.850  &   PIMC\_A3377 \\
 50.1743 &    4041916 &     856514.5 &   3610.6 &     569.466 &  2.272  &   PIMC\_A3376 \\
 50.1743 &    8083831 &    1808604.3 &   4427.3 &    1194.311 &  2.786  &   PIMC\_A3375 \\
 50.1743 &   16167663 &    3717607.4 &   4748.5 &    2413.995 &  2.991  &   PIMC\_A3374 \\
 50.1743 &   32335325 &    7494789.2 &   7111.9 &    4803.281 &  4.481  &   PIMC\_A3373 \\
 50.1743 &   64670651 &   15097727.3 &   9275.7 &    9596.045 &  5.833  &   PIMC\_A3372 \\
 50.1743 &  129341301 &   30227397.3 &  12043.7 &   19120.537 &  7.599  &   PIMC\_A3371 \\
 50.1743 &  258682602 &   60498356.0 &  19034.1 &   38169.777 & 11.969  &   PIMC\_A3370 \\
 50.1743 &  517365204 &  121037580.3 &  17009.5 &   76260.744 & 10.735  &   PIMC\_A3369 \\
  2.5090 &       2000 &          9.3 &      0.8 &       0.231 &  0.001   & PBE\_ONCV\_1.07 \\
  2.5090 &       6736 &         34.8 &      2.7 &       0.517 &  0.002   & PBE\_ONCV\_1.07 \\
  2.5090 &      10000 &         59.7 &      3.9 &       0.658 &  0.004   & PBE\_ONCV\_1.07 \\
  2.5090 &      35001 &        176.0 &      5.1 &       1.921 &  0.006   & PBE\_ONCV\_1.07 \\
  2.5090 &     126313 &        714.1 &      8.2 &       8.732 &  0.018   & PBE\_ONCV\_1.07 \\
  2.5090 &     673653 &       4948.3 &     34.5 &      78.542 &  0.149   & PBE\_ONCV\_2.06 \\
  2.5090 &     842066 &       6686.7 &     22.2 &     111.475 &  0.101   & PBE\_ONCV\_2.06 \\
  2.5090 &    1010479 &       8472.6 &     24.3 &     146.350 &  0.157   & PBE\_ONCV\_2.06 \\
  2.5090 &    1347305 &      12374.3 &     15.2 &     214.048 &  0.098   & PBE\_ONCV\_2.06 \\
  5.0170 &       6736 &        438.2 &      4.8 &       0.932 &  0.002   & PBE\_ONCV\_1.07 \\
  5.0170 &      10000 &        488.2 &      6.3 &       1.113 &  0.004   & PBE\_ONCV\_1.07 \\
  5.0170 &      35001 &        728.7 &     13.3 &       2.308 &  0.011   & PBE\_ONCV\_1.07 \\
  5.0170 &     126313 &       1821.7 &     20.6 &       8.701 &  0.026   & PBE\_ONCV\_1.07 \\
  5.0170 &     350013 &       4898.4 &     12.6 &      28.724 &  0.043   & PBE\_ONCV\_2.06 \\
  5.0170 &     673653 &       9984.3 &     10.1 &      71.185 &  0.055   & PBE\_ONCV\_2.06 \\
  5.0170 &     842066 &      12831.4 &     12.2 &     100.502 &  0.057   & PBE\_ONCV\_2.06 \\
  5.0170 &    1010479 &      15990.7 &     33.0 &     130.330 &  0.106   & PBE\_ONCV\_2.06 \\
  5.0170 &    1347305 &      23939.0 &      6.7 &     193.193 &  0.102   & PBE\_ONCV\_2.06 \\
  7.5260 &       6736 &       1316.7 &      7.0 &       1.894 &  0.004   & PBE\_ONCV\_1.07 \\
  7.5260 &      10000 &       1376.0 &      9.2 &       2.084 &  0.005   & PBE\_ONCV\_1.07 \\
  7.5260 &      35001 &       1770.7 &     14.1 &       3.298 &  0.009   & PBE\_ONCV\_1.07 \\
  7.5260 &     126313 &       3330.8 &     47.2 &       9.306 &  0.039   & PBE\_ONCV\_1.07 \\
  7.5260 &     350013 &       7905.6 &     28.0 &      28.870 &  0.066   & PBE\_ONCV\_2.06 \\
  7.5260 &     673653 &      15539.5 &     23.1 &      67.735 &  0.060   & PBE\_ONCV\_2.06 \\
  7.5260 &     842066 &      20140.4 &     23.1 &      93.497 &  0.079   & PBE\_ONCV\_2.06 \\
  7.5260 &    1010479 &      24590.6 &     16.9 &     119.636 &  0.104   & PBE\_ONCV\_2.06 \\
  7.5260 &    1347305 &      34532.3 &     22.9 &     180.619 &  0.133   & PBE\_ONCV\_2.06 \\
 10.0350 &       6736 &       2690.7 &     10.4 &       3.140 &  0.003   & PBE\_ONCV\_1.07 \\
 10.0350 &      10000 &       2751.2 &      4.1 &       3.308 &  0.003   & PBE\_ONCV\_1.07 \\
 10.0350 &      35001 &       3297.3 &     27.5 &       4.589 &  0.015   & PBE\_ONCV\_1.07 \\
 10.0350 &     126313 &       5327.5 &     46.0 &      10.417 &  0.034   & PBE\_ONCV\_1.07 \\
 10.0350 &     350013 &      11301.0 &     49.0 &      29.423 &  0.068   & PBE\_ONCV\_2.06 \\
 10.0350 &     673653 &      21476.6 &     91.1 &      66.968 &  0.216   & PBE\_ONCV\_2.06 \\
 10.0350 &     842066 &      27203.8 &     42.3 &      90.131 &  0.157   & PBE\_ONCV\_2.06 \\
 10.0350 &    1010479 &      33457.0 &     74.1 &     117.018 &  0.187   & PBE\_ONCV\_2.06 \\
 10.0350 &    1347305 &      46468.1 &     63.8 &     174.733 &  0.148   & PBE\_ONCV\_2.06 \\
 11.2890 &       6736 &       3543.6 &      8.2 &       3.832 &  0.004   & PBE\_ONCV\_1.07 \\
 11.2890 &      10000 &       3634.3 &     10.9 &       4.017 &  0.005   & PBE\_ONCV\_1.07 \\
 11.2890 &      35001 &       4206.0 &     31.0 &       5.246 &  0.016   & PBE\_ONCV\_1.07 \\
 11.2890 &     126313 &       6486.6 &     56.3 &      11.048 &  0.043   & PBE\_ONCV\_1.07 \\
 11.2890 &     350013 &      13071.0 &     35.4 &      29.683 &  0.052   & PBE\_ONCV\_2.06 \\
 11.2890 &     673653 &      24818.9 &    141.9 &      67.657 &  0.034   & PBE\_ONCV\_2.06 \\
 11.2890 &     842066 &      30471.3 &     16.0 &      88.406 &  0.011   & PBE\_ONCV\_2.06 \\
 11.2890 &    1010479 &      38382.2 &     67.2 &     116.519 &  0.023   & PBE\_ONCV\_2.06 \\
 11.2890 &    1347305 &      52575.8 &     46.7 &     172.974 &  0.015   & PBE\_ONCV\_2.06 \\
 12.5440 &       6736 &       4519.2 &      2.9 &       4.543 &  0.002   & PBE\_ONCV\_2.06 \\
 12.5440 &      10000 &       4663.0 &      1.8 &       4.767 &  0.001   & PBE\_ONCV\_2.06 \\
 12.5440 &      35001 &       5302.7 &     16.2 &       6.002 &  0.012   & PBE\_ONCV\_2.06 \\
 12.5440 &     126313 &       7840.6 &     17.6 &      11.940 &  0.025   & PBE\_ONCV\_2.06 \\
 12.5440 &     350013 &      15317.2 &     77.9 &      30.819 &  0.105   & PBE\_ONCV\_2.06 \\
 12.5440 &     673653 &      26264.0 &     19.6 &      64.244 &  0.071   & PBE\_ONCV\_2.06 \\
 12.5440 &     842066 &      34878.4 &     81.5 &      89.036 &  0.069   & PBE\_ONCV\_2.06 \\
 12.5440 &    1010479 &      42918.5 &     22.6 &     115.451 &  0.023   & PBE\_ONCV\_2.06 \\
 12.5440 &    1347305 &      58855.8 &     43.6 &     167.049 &  0.067   & PBE\_ONCV\_2.06 \\
 15.0520 &       2000 &       6618.1 &      1.1 &       5.769 &  0.002   & PBE\_ONCV\_2.06 \\
 15.0520 &       6736 &       6809.5 &      2.6 &       6.074 &  0.003   & PBE\_ONCV\_2.06 \\
 15.0520 &      10000 &       6925.4 &      6.0 &       6.269 &  0.004   & PBE\_ONCV\_2.06 \\
 15.0520 &      35001 &       7623.3 &     21.8 &       7.432 &  0.016   & PBE\_ONCV\_2.06 \\
 15.0520 &     126313 &      10488.1 &     57.0 &      12.961 &  0.056   & PBE\_ONCV\_2.06 \\
 15.0520 &     350013 &      19377.3 &    101.4 &      31.455 &  0.103   & PBE\_ONCV\_2.06 \\
 15.0520 &     673653 &      33914.0 &     30.5 &      65.918 &  0.090   & PBE\_ONCV\_2.06 \\
 15.0520 &     842066 &      41941.1 &     79.4 &      92.213 &  0.132   & PBE\_ONCV\_2.06 \\
 15.0520 &    1010479 &      51904.3 &     97.5 &     113.040 &  0.125   & PBE\_ONCV\_2.06 \\
 15.0520 &    1347305 &      72318.8 &    342.6 &     169.570 &  0.169   & PBE\_ONCV\_2.06 \\
 17.5610 &      10000 &       9638.9 &      6.7 &       7.734 &  0.003   & PBE\_ONCV\_2.06 \\
 17.5610 &      35001 &      10408.7 &      6.4 &       8.958 &  0.008   & PBE\_ONCV\_2.06 \\
 17.5610 &     126313 &      13972.8 &     24.0 &      14.757 &  0.032   & PBE\_ONCV\_2.06 \\
 17.5610 &     350013 &      24066.7 &    124.8 &      33.087 &  0.124   & PBE\_ONCV\_2.06 \\
 17.5610 &     673653 &      40863.3 &    152.1 &      66.182 &  0.149   & PBE\_ONCV\_2.06 \\
 17.5610 &     842066 &      50853.8 &    154.1 &      88.308 &  0.156   & PBE\_ONCV\_2.06 \\
 17.5610 &    1010479 &      62119.5 &    333.4 &     116.191 &  0.141   & PBE\_ONCV\_2.06 \\
 17.5610 &    1347305 &      84179.8 &     44.1 &     163.040 &  0.124   & PBE\_ONCV\_2.06 \\
 20.0700 &      10000 &      12811.8 &     11.2 &       9.369 &  0.007   & PBE\_ONCV\_2.06 \\
 20.0700 &      35001 &      13915.9 &      8.6 &      10.773 &  0.005   & PBE\_ONCV\_2.06 \\
 20.0700 &     126313 &      17586.3 &     65.4 &      16.123 &  0.051   & PBE\_ONCV\_2.06 \\
 20.0700 &     350013 &      28877.3 &    115.3 &      34.082 &  0.093   & PBE\_ONCV\_2.06 \\
 20.0700 &     673653 &      48008.1 &    224.8 &      67.112 &  0.030   & PBE\_ONCV\_2.06 \\
 20.0700 &     842066 &      59253.3 &    301.9 &      91.678 &  0.024   & PBE\_ONCV\_2.06 \\
 20.0700 &    1010479 &      70396.8 &     89.0 &     111.236 &  0.066   & PBE\_ONCV\_2.06 \\
 20.0700 &    1347305 &      97864.6 &    171.5 &     161.321 &  0.097   & PBE\_ONCV\_2.06 \\
 25.0870 &      10000 &      20383.0 &      5.2 &      12.629 &  0.003   & PBE\_ONCV\_2.06 \\
 25.0870 &      35001 &      21740.1 &     28.1 &      14.131 &  0.014   & PBE\_ONCV\_2.06 \\
 25.0870 &     126313 &      26153.1 &     50.1 &      19.250 &  0.036   & PBE\_ONCV\_2.06 \\
 25.0870 &     350013 &      40240.6 &    173.9 &      37.224 &  0.116   & PBE\_ONCV\_2.06 \\
 25.0870 &     673653 &      62748.5 &    133.9 &      67.642 &  0.119   & PBE\_ONCV\_2.06 \\
 25.0870 &     842066 &      76860.4 &    451.1 &      91.864 &  0.114   & PBE\_ONCV\_2.06 \\
 25.0870 &    1010479 &      92273.8 &    551.9 &     110.436 &  0.141   & PBE\_ONCV\_2.06 \\
 25.0870 &    1347305 &     124901.1 &    564.6 &     161.060 &  0.127   & PBE\_ONCV\_2.06 \\
 30.1050 &      10000 &      29636.4 &     12.5 &      15.982 &  0.004   & PBE\_ONCV\_2.06 \\
 30.1050 &      35001 &      31097.4 &     41.0 &      17.386 &  0.014   & PBE\_ONCV\_2.06 \\
 30.1050 &     126313 &      36480.6 &     81.2 &      22.816 &  0.043   & PBE\_ONCV\_2.06 \\
 30.1050 &     350013 &      52548.5 &    157.1 &      39.666 &  0.099   & PBE\_ONCV\_2.06 \\
 30.1050 &     673653 &      79783.9 &    529.2 &      70.934 &  0.106   & PBE\_ONCV\_2.06 \\
 30.1050 &     842066 &      98505.0 &    204.4 &      90.862 &  0.108   & PBE\_ONCV\_2.06 \\
 30.1050 &    1010479 &     114099.0 &    653.8 &     113.600 &  0.100   & PBE\_ONCV\_2.06 \\
 30.1050 &    1347305 &     151123.5 &    518.9 &     159.254 &  0.082   & PBE\_ONCV\_2.06 \\
 37.6310 &      10000 &      46591.7 &     14.1 &      21.192 &  0.005   & PBE\_ONCV\_2.06 \\
 37.6310 &      35001 &      48478.9 &     56.0 &      22.750 &  0.020   & PBE\_ONCV\_2.06 \\
 37.6310 &     126313 &      54509.5 &    135.1 &      27.698 &  0.071   & PBE\_ONCV\_2.06 \\
 37.6310 &     350013 &      74028.7 &    839.7 &      43.372 &  0.081   & PBE\_ONCV\_2.06 \\
 37.6310 &     673653 &     108684.7 &    484.0 &      74.384 &  0.042   & PBE\_ONCV\_2.06 \\
 37.6310 &     842066 &     127839.8 &    630.1 &      93.539 &  0.050   & PBE\_ONCV\_2.06 \\
 37.6310 &    1010479 &     152358.0 &    600.1 &     113.126 &  0.070   & PBE\_ONCV\_2.06 \\
 37.6310 &    1347305 &     200300.2 &   1428.8 &     160.424 &  0.047   & PBE\_ONCV\_2.06 \\
 50.1750 &      10000 &      82106.4 &     25.8 &      29.732 &  0.008   & PBE\_ONCV\_2.06 \\
 50.1750 &      35001 &      84498.8 &     67.5 &      31.221 &  0.026   & PBE\_ONCV\_2.06 \\
 50.1750 &     126313 &      91888.0 &    151.8 &      35.933 &  0.055   & PBE\_ONCV\_2.06 \\
 50.1750 &     350013 &     117311.6 &    756.2 &      59.691 &  0.298   & PBE\_ONCV\_2.06 \\
 50.1750 &     673653 &     160348.0 &    612.3 &      83.482 &  0.161   & PBE\_ONCV\_2.06 \\
 50.1750 &     842066 &     188902.7 &   1262.9 &      99.851 &  0.415   & PBE\_ONCV\_2.06 \\
 50.1750 &    1010479 &     213986.7 &   1638.1 &     117.712 &  0.518   & PBE\_ONCV\_2.06 \\
 50.1750 &    1347305 &     273743.7 &   1461.5 &     159.957 &  0.224   & PBE\_ONCV\_2.06 \\
\end{longtable*}

\end{document}